\documentclass[12pt,letterpaper]{article}
\usepackage[utf8]{inputenc}

\usepackage{graphicx,array}
\usepackage{url}
\usepackage{color}
\usepackage{latexsym}
\usepackage{amsthm}
\usepackage{amsmath}
\usepackage{amssymb}
\usepackage{amsfonts}
\usepackage[numbers,sort&compress]{natbib}
\usepackage{bm}
\usepackage{slashed}
\usepackage{mathrsfs}
\usepackage{enumerate}
\usepackage{tikz}
\usepackage{siunitx}
\usepackage{mdframed}
\usepackage{setspace}  
\usepackage{esvect}
\usepackage{caption}

\usepackage{tcolorbox}%

\usepackage{hyperref} 
\hypersetup{
    colorlinks=true,       
    linkcolor=red,          
    citecolor=blue,        
    filecolor=magenta,      
    urlcolor=blue           
}
\usepackage[all]{hypcap} 
\usepackage{multirow}
\usepackage{multicol}
\usepackage{makecell}

\usepackage{natbib}
\newcommand{\nc}{\newcommand}  
\newcommand{\mc}{\mathcal}

\newcommand{\uu}{\;\!}

\nc{\beq}{\begin{equation}}
\nc{\eeq}{\end{equation}}
\nc{\beqa}{\begin{eqnarray}}  
\nc{\eeqa}{\end{eqnarray}}  
\nc{\bit}{\begin{itemize}}  
\nc{\eit}{\end{itemize}}  

\newcommand{\eg}{{\it e.g.}}
\newcommand{\ie}{{\it i.e.}}

\newcommand{\lcdm}{\Lambda\text{CDM}}
\newcommand{\xdw}{x_{\rm DW}}
\newcommand{\at}{a_t}
\newcommand{\zt}{z_t}
\newcommand{\fa}{f_{\rm DM}} 
\newcommand{\DV}{\Delta V}
\newcommand{\txpy}{t_{\langle x,y \rangle}}
\newcommand{\txmy}{t_{x,y}}
\newcommand{\txyn}{t_{\langle x,y \rangle,n}}

\title{{\bf Late-time Quantum Vacuum Decay and its Cosmological Implications}
}
\author{\large Yang Bai$^{\,\circ\natural}$, Sida Lu$^{\,\dagger\footnote{Corresponding author}}$\,, and Nicholas Orlofsky$^{\, \diamond}$}
\date{\small \it 
$^\dagger$School of Physics and Astronomy, Sun Yat-sen University (Zhuhai Campus), Zhuhai 519082, China\\
$^\circ$Department of Physics, University of Wisconsin-Madison, Madison, WI 53706, USA\\
$^\natural$HEP Division, Argonne National Laboratory, Argonne, IL 60439, USA \\
$^\diamond$Institute of Theoretical Physics, Faculty of Physics, University of Warsaw, ul. Pasteura 5, PL-02-093 Warsaw, Poland \\}

\begin{document}

\maketitle

\setlength{\parskip}{0.2ex}

\begin{abstract} 
The existence of a landscape of metastable vacua raises the possibility that our Universe may have undergone quantum vacuum decay at late times. This work explores how such a transition can be tested with cosmological observables, focusing on precision distance measurements and cosmic microwave background (CMB) anisotropies. A set of phenomenological models is constructed in which late-time quantum tunneling changes the vacuum energy and may convert a subcomponent of dark matter into dark radiation, possibly accompanied by domain-wall production. The resulting expansion histories are compared with DESI DR2 baryon acoustic oscillation data; supernova distance measurements from DES-Dovekie, Pantheon+, and Union3; and a compressed CMB likelihood. 
For quantum-tunneling models, current cosmological distance measurements still allow a 50\% decrease in the total vacuum energy for a transition redshift $z_t<1$.  
The model with dark-matter conversion and domain-wall production provides a good fit to resolve the tension between cosmological observables and the $\Lambda$CDM model, with a preferred transition around $z_t \sim 7$ and about 10\% of dark matter participating in the transition. 
Additionally, CMB anisotropy constraints from bubble nucleation and the associated domain-wall network are derived and shown to strongly restrict slow or sparse late transitions. Applied to the minimal quantum-tunneling model, these constraints allow an $\mathcal{O}(10\%)$ decrease in the total vacuum energy for a transition redshift $z_t$ of order unity. For nonminimal models, dark-matter-density-dependent tunneling can proceed rapidly enough to evade such bounds.  These results demonstrate that late-time quantum vacuum decay is a testable cosmological phenomenon and provide a concrete observational handle on metastable-vacuum physics motivated by landscape scenarios.

\end{abstract}

\thispagestyle{empty}  
\newpage   
\setcounter{page}{1}  

\begingroup
\hypersetup{linkcolor=black,linktocpage}
\tableofcontents
\endgroup

\newpage

\section{Introduction}
\label{sec:intro}

Vacua and transitions between them have long been intriguing topics across all areas of physics. Among the most fundamental ideas related to these topics is the string landscape~\cite{Bousso:2000xa, Giddings:2001yu, Kachru:2003aw, Susskind:2003kw}, which, in combination with the anthropic principle~\cite{Weinberg:1987dv}, provides a possible explanation for the severe hierarchy between the vacuum energy of the Universe, namely the cosmological constant (CC), and the Planck scale (see, \eg, \cite{Polchinski:2006gy,Brennan:2017rbf,Agmon:2022thq} for reviews).
The current understanding suggests that there may exist as many as $10^{500}$ local minima~\cite{Denef:2004ze} with different vacuum energies, including some near zero CC.
The possible values of the CC (as well as other important quantities~\cite{Arkani-Hamed:2005zuc}) are sampled by these metastable vacua, and our observed Universe is among those compatible with successful structure formation, a suitable amount of baryon asymmetry, and other properties necessary for intelligent observers to exist.
Of course, multiple vacua can arise even in much simpler theories containing minimal extra scalar degrees of freedom. Even the Standard Model coupled to gravity on its own contains a landscape of lower-dimensional vacua \cite{Arkani-Hamed:2007ryu}, and the current vacuum is thought to be metastable if no new physics modifies the running of the Higgs quartic coupling~\cite{Isidori:2001bm}.

A plethora of studies have been devoted to understanding the landscape, including the construction, counting, and classification of vacua and the underlying manifolds, as well as statistical treatments of the vacua~\cite{Arkani-Hamed:2005zuc,Greene:2013ida,Wang:2015rel,Gendler:2023ujl,McAllister:2024lnt}.
In addition to these efforts, it is also important to identify possible low-energy tests of the landscape picture, \eg, searches for supersymmetric theories emergent from the string landscape or for light moduli naturally arising in string theory~\cite{DAgnolo:2025cxb,Baer:2025zqt,Baer:2023ech,Gendler:2024adn}.
Given the intrinsically cosmological nature of this idea, it would be particularly interesting to test or constrain the landscape picture using cosmological observations.
Since vacua, including that of our own Universe, are generically metastable, it is natural to expect quantum tunneling transitions (QTs) between them.
Specifically, it has been argued that the tunneling rate depends sensitively on the number of moduli involved in the landscape~\cite{Greene:2013ida,Dine:2015ioa}, although this conclusion can be realization dependent; see, \eg, \cite{Masoumi:2016eqo,Wang:2015rel}.
Possible tests or constraints on these QTs may therefore provide further clues about the fundamental theories underlying the landscape.

The chief concern of this work is QTs occurring in the late Universe at redshifts of $\mc{O}(1)$. Such a scenario could arise rather generically in the string landscape framework: some metastable vacua are expected to have vacuum energies and field-space locations not far from the observed CC, making a QT that takes place closer to the present epoch more plausible. Late-time vacuum transitions have historically received less attention than early-Universe phase transitions (PTs)~\cite{Kamionkowski:1993fg,Caprini:2015zlo,Bringmann:2023opz,Ghosh:2023aum,Salvio:2023ynn,Winkler:2024olr}, perhaps in part because they are not expected to produce a stochastic gravitational-wave background detectable by current experiments~\cite{LISACosmologyWorkingGroup:2022jok,Luo:2025ewp,Hu:2017mde,NANOGrav:2023gor} (however, see~\cite{Patwardhan:2014iha,Krauss:2007rx} for earlier discussions). 
In this work, we explore alternative avenues for probing late-Universe QTs using cosmological observables, including the expansion history which can probe changes in the equation-of-state parameter, and anisotropies produced by the QT itself or by its relics. 
We now briefly introduce these possible probes of late-Universe QTs in turn.

The expansion history is one of the most direct probes of the changes brought by a QT. When a QT takes place, the vacuum energy of the Universe changes, and part of this change may be converted into other forms of energy, such as dark radiation (DR). In addition, it has been shown that a PT may assist the formation of primordial black holes~\cite{Lewicki:2024ghw,Liu:2021svg,Kanemura:2024pae}, create topological defects such as domain walls~\cite{Kolb:1992uu,Bai:2023cqj,Bai:2025qch}, and have interesting interplays with the mass and relic abundance of dark matter~\cite{Baker:2019ndr,Wong:2023qon,Bai:2022kxq}.
The latter two possibilities are considered in this work\footnote{Primordial black hole formation and gravitational waves from QTs were discussed recently in~\cite{An:2026hiq}, although in that case the transitions occur prior to matter-radiation equality.}, where we construct a series of toy models to capture the essential features of what may occur during a late QT.
In order of increasing complexity, these models are as follows:
\begin{itemize}
    \item QT: A simple vacuum PT (\ie, quantum tunneling without any finite-temperature or density effects) in which the entire change in vacuum energy is converted into DR. The tunneling rate in this case is taken to be constant in time. (A variant of this model, QT+DW, in which domain walls (DWs) are also formed, is briefly mentioned but not studied in detail due to stringent constraints.)
    \item QT+DM: Similar to the QT model, but allowing for the possibility that a fraction of dark matter (DM) is converted into DR after the QT. This may occur if there is a nontrivial coupling between the scalar field undergoing the transition and the DM mass term, especially if the final vacuum expectation value of the scalar field after the transition vanishes. As we will see, this coupling to DM can induce a DM-density dependence in the tunneling rate.
    \item QT+DM+DW: This model includes all of the above ingredients, while also allowing domain walls to form at the end of the transition, assuming a $\mathbb{Z}_2$ symmetry for the scalar field undergoing the transition. Note that unlike early-universe PTs, DW formation does not pose a cosmological overclosure concern in this case because the QT occurs sufficiently late.
\end{itemize}

To assess their impact on the expansion history, these models are fitted to the recent DESI DR2 baryon acoustic oscillation (BAO) measurements~\cite{DESI:2025zgx}, combined with cosmic microwave background (CMB) anisotropy data and the DES-Dovekie~\cite{DES:2025sig}, Pantheon+~\cite{Scolnic:2021amr,Brout:2022vxf}, and Union3~\cite{Rubin:2023jdq} supernova (SN) distance measurements.
Fits to this data combination have been shown to generically exhibit a discrepancy with the standard $\lcdm$ model at the 3-$\sigma$ level~\cite{DESI:2025zgx}. Using the CPL (Chevallier-Polarski-Linder) parametrization of the dark energy equation of state (EoS) as a function of scale factor $w(a)=w_0+w_a (1-a)$~\cite{Chevallier:2000qy,Linder:2002et}, the data suggest that dark energy may evolve from a ``phantom'' phase with $w<-1$ to one with $w>-1$ at recent times. Possible interpretations arise in the context of modified gravity, nonstandard dark energy, or a nonstandard matter sector~\cite{Braglia:2025gdo,LaPenna:2026avs,An:2025vfz,Khoury:2025txd,Petri:2025swg,Chen:2025ywv,Silva:2025hxw,Tsujikawa:2026xqm,Nojiri:2025uew,SanchezLopez:2025uzw,Lee:2025pzo,Li:2025eqh,Cai:2025mas,Mirpoorian:2025rfp,Chen:2025wwn}.

We find that the QT+DM+DW model provides a significantly better fit to these data than the $\Lambda$CDM model, performing comparably to or better than the CPL model. Although the QT+DM+DW model appears more complicated at first glance than the CPL model because it has two additional parameters, the CPL model is merely an effective parametrization, whereas the QT+DM+DW model is based on an underlying physical framework. On the other hand, the QT and QT+DM models do not significantly improve the fit, although meaningful bounds can still be placed on their free parameters. 
Specifically, we find that SN distance and BAO measurements still allow a 50\% decrease in the vacuum energy if the transition occurs and completes at a redshift below one, whereas CMB anisotropy constraints are more stringent, allowing only an $\mathcal{O}(10\%)$ or smaller decrease for a QT completed at a redshift of $\mathcal{O}(1)$.

Meanwhile, it has been shown that fluctuations generated by late transitions may produce curvature perturbations that leave imprints on the CMB anisotropy spectrum~\cite{Elor:2023xbz,Koren:2025ymq,Lewicki:2024ghw,Franciolini:2025ztf}, similar to the Integrated Sachs--Wolfe (ISW) effect~\cite{Sachs:1967er}. These previous studies focused on thermal PTs with time-dependent nucleation rates. We apply their results to our models and extend them in two ways. First, we derive the corresponding constraints for vacuum PTs whose nucleation rates are time independent. Second, for models that also produce DWs, we estimate the associated anisotropy bounds arising from the DWs themselves. Both types of anisotropy constraints can impose rather stringent limits on slowly evolving or time-independent nucleation rates. On the other hand, QTs with more rapidly evolving nucleation rates can generally evade all of these bounds.

The remainder of this work is organized as follows. We first introduce the three QT-involved models outlined above in Sec.~\ref{sec:model}.
Then in Sec.~\ref{sec:cosmology}, we examine the possible influences of these models on cosmology. We begin by examining how the cosmic background evolution is modified by these three models. We then fit the models to the combination of DESI DR2 BAO data, SN distance moduli measurements, and the CMB. After that, we examine the possible imprint of the late QT on the CMB anisotropy, and draw constraints from the measured power spectrum. Finally, we conclude in Sec.~\ref{sec:discussion}. Details regarding the data visualization in Sec.~\ref{sec:background_evo} are provided in Appendix~\ref{app:binning},
while details of some calculations for the CMB anisotropies in Sec.~\ref{sec:ISW-vacuum} are given in Appendices~\ref{sec:bubble-calc} and~\ref{app:power-spectra}.

\begin{figure}[t!]
    \centering
    \includegraphics[width=0.6\linewidth]{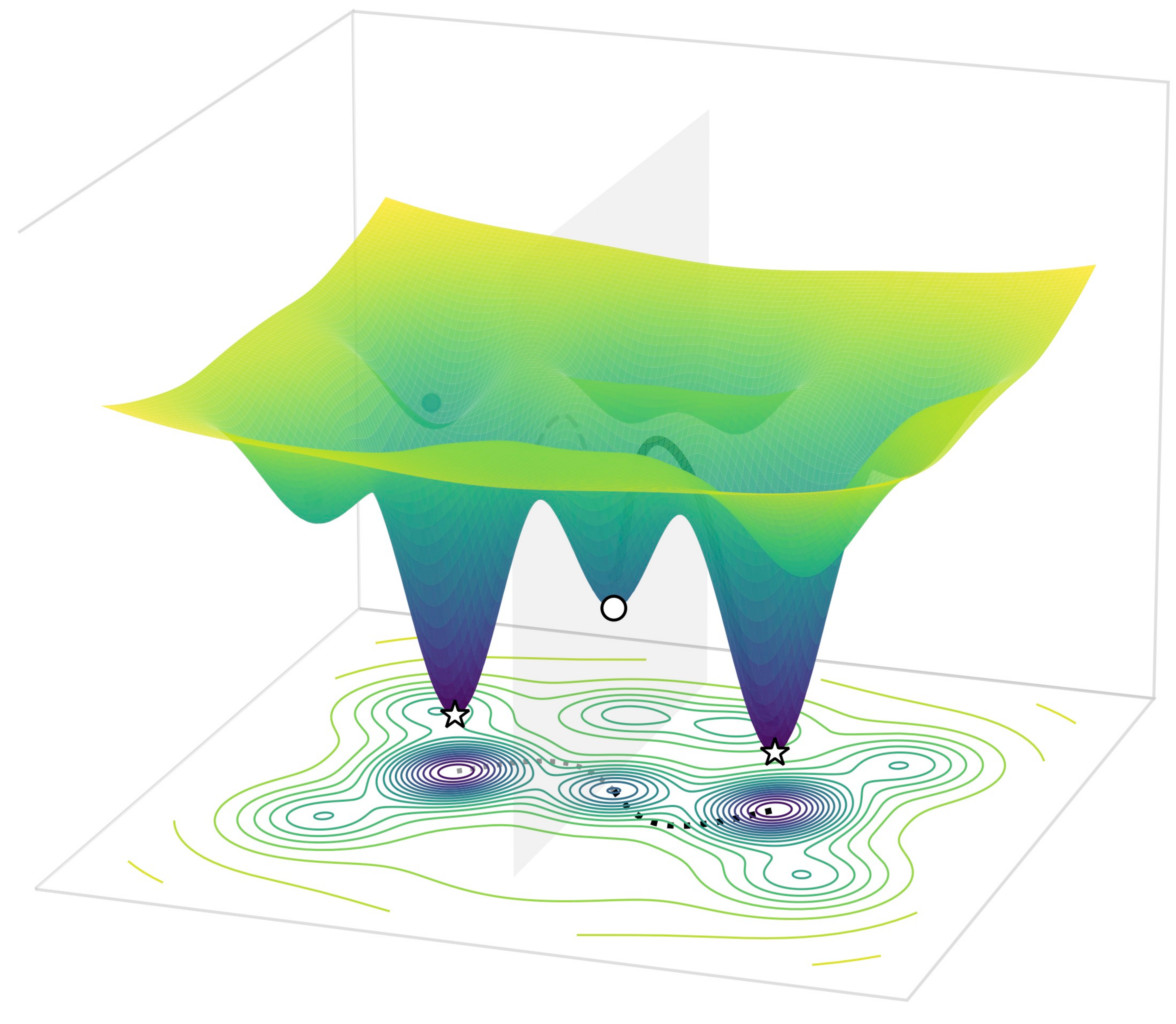}   
    \caption{Schematic illustration of quantum tunneling in vacuum decay. The circle denotes a nearby metastable vacuum that tunnels to the current-universe vacuum, marked by the five-pointed stars. For a discrete $\mathcal{Z}_2$ symmetry centered on the metastable vacuum, domain walls may be generated during the vacuum decay process. The base plane displays contours of equal potential, and the dotted curve represents a possible tunneling trajectory. } 
    \label{fig:landscape}
\end{figure}

\label{sec:DESI-data}

\section{Models with late-time quantum vacuum decay}
\label{sec:model}

The Universe may contain a landscape of local minima determined by many field degrees of freedom as schematically illustrated in Fig.~\ref{fig:landscape}. For the purposes of this work, it is helpful to focus only on a small set of fields that are relevant to the purported late-universe transition. Therefore, various toy models will be presented, which can be thought of as effective theories near the local minima in question, but that could be embedded in a larger theory.

In this section, we present three such models in order of increasing complexity. These will later be used to examine the potential cosmological consequences for late-time vacuum decays from a metastable vacuum. The first and simplest model is dubbed ``Quantum tunneling" (QT). Since the order-parameter field could couple to the dark sector, the ``Quantum tunneling + dark matter" (QT+DM) model explores scenarios with the DM finite density playing a nontrivial role in modifying the vacuum decay rates. Furthermore, the simple scalar-field potential used in these models may contain a discrete $\mathbb{Z}_2$ symmetry, which could be spontaneously broken during the quantum tunneling process and lead to the late-time production of DWs. This final possibility is explored in the ``Quantum tunneling + dark matter + domain wall" (QT+DM+DW) model.

The case of quantum tunneling + domain walls (QT+DW) could also arise from the QT model presented in Sec.~\ref{sec:model-QT}. We do not go into detail on the QT+DW case because, as we will see in Sec.~\ref{sec:ISW}, it is severely constrained by CMB anisotropies arising from the DWs.

\subsection{Quantum tunneling (QT)}
\label{sec:model-QT}

Consider a model with two scalar fields having the tree level potential
\beqa
V_{\mathbb{Z}_2^{\phi S}} = \frac{\lambda}{4}\,(\phi^2-v^2)^2 + \frac{\lambda}{4}\,(S^2 - v^2)^2 + \frac{\kappa}{2}\,\phi^2\,S^2 - \frac{\lambda}{4} v^4 + V_0 \,,
\label{eq:Z2-symmetric-potential}
\eeqa
with a discrete matter interchange symmetry: 
\beqa
\mathbb{Z}_2^{\phi S}:\qquad  \phi \leftrightarrow S ~.
\eeqa
Here, $\lambda, \kappa > 0$ for the potential to be bounded from below. There are two additional $\mathbb{Z}_2$ symmetries: $\mathbb{Z}^\phi_2$ for $\phi \rightarrow - \phi$ and $\mathbb{Z}^S_2$ for $S \rightarrow -S$.\footnote{For the QT and QT+DM models, $\phi$ and $S$ need not be real. They could instead have some larger symmetry group that enforces that they only enter the potential in integer powers of $|\phi|^2$ or $|S|^2$. Indeed, in the QT+DM and QT+DM+DW models, a larger symmetry group for $S$ is explicitly used. Only the QT+DM+DW model requires $\phi$ to be real.} 
For
\beqa
\kappa > \lambda ~,
\eeqa
this potential admits four degenerate local vacua: $(\langle \phi \rangle, \langle S \rangle) = (\pm v, 0), (0, \pm v)$. We have introduced the term $-\frac{\lambda}{4} v^4$ in \eqref{eq:Z2-symmetric-potential} to have the CC at the minima be $V_0$. 

To lift some of the local vacua, one could introduce a discrete symmetry breaking operator. Without loss of generality, we introduce the following $\mathbb{Z}_2^{\phi S}$-breaking operator
\beqa
\label{eq:V-phiS-breaking}
V_{\slashed{\mathbb{Z}}_2^{\phi S}} = \frac{\xi}{4} S^4 ~,
\eeqa
with $\xi \ll \lambda, \kappa$. 
The resulting vacua are 
\beqa
&& \mbox{Vac}_\phi: (\pm v, 0)\,, \quad \mbox{with} \quad V_\phi =  V_0 ~,\\
&& \mbox{Vac}_S: (0, \pm v)\,, \quad \mbox{with} \quad V^0_S = V_0 + \frac{\xi}{4}\,v^4~,
\label{eq:V_S^0}
\eeqa
where we have ignored the small shift of the vacuum expectation value (VEV) of $S$ due to the $\xi$ operator. For $\xi > 0\,(<0)$, Vac$_\phi$ (Vac$_S$) is the global minimum of the model. We take $\xi > 0$, so that quantum tunneling from the metastable vacuum to the global vacuum proceeds as
\beqa
\label{eq:tunneling}
 \mbox{Vac}_S \xrightarrow{\text{tunneling}}  \mbox{Vac}_\phi ~.
\eeqa
This tunneling allows the possibility for DW formation. A further explicit breaking of the $\mathbb{Z}_2^{\phi}$ symmetry could prevent DWs. In the following, we assume this explicit breaking (if present) to be small enough to have a negligible effect on tunneling rate calculations.

Depending on physics in the very early universe, the visible universe could sit in the metastable vacuum Vac$_S$. If the lifetime of this vacuum is longer than the age of universe, one may not need to worry about its metastability. If the lifetime is relatively short, say shorter than the Big Bang nucleosynthesis (BBN) time, the only sign today of such an early QT may be in the form of relics like stochastic gravitational waves or dark radiation. 
For a lifetime between BBN and the current era, the quantum tunneling of the metastable vacuum can have nontrivial effects on cosmological observables, which will be the focus of this study. 

For the very degenerate vacua case, the thin-wall approximation provides a good estimation for the lifetime of the metastable vacuum. Following Coleman \cite{Coleman:1977py,Callan:1977pt}, the four-dimensional Euclidean action for a vacuum PT in the thin-wall approximation is
\begin{equation}
    S_4=\frac{27\pi^2S^4_1}{2\,\DV^3}\,,
    \label{eq:thin-wall-S4}
\end{equation}
where
\beqa\label{eq:S1_2_field}
S_1&=&\int dr\left[\frac{1}{2}\left(\frac{d \phi}{dr}\right)^2+\frac{1}{2}\left(\frac{d S}{dr}\right)^2 + \left[V(\phi, S)- V(\phi^*, S^*)\right] \right] \nonumber \\
&=& \int_{0}^{\phi^*} d\phi \sqrt{1 + \left(\frac{dS}{d\phi}\right)^2}\,\sqrt{2\,[V(\phi, S)-V(\phi^*, S^*)]}\,,
\eeqa
and $\DV = \frac{\xi}{4}\,v^4$ is the vacuum energy difference between the true and false vacuum. The field value $(\phi^*, S^*)$ is the release point on the true vacuum side of the potential and satisfies $V(0, \pm v)=V(\phi^*, S^*)$. For generic values of $\lambda, \kappa$, one can numerically solve the differential equations to obtain the bounce profile as well as $S_1$. 

\begin{figure}[t!]
    \centering
    \includegraphics[width=0.48\linewidth]{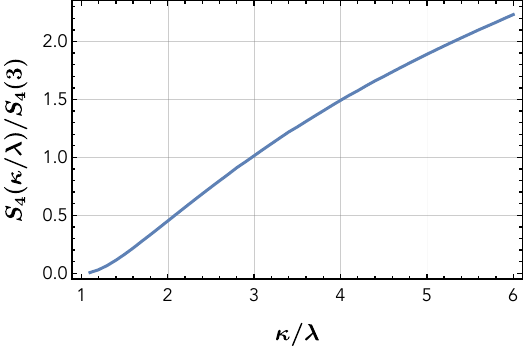}  
    \caption{The numerically obtained four-dimensional Euclidean action $S_4$ as a function of $\kappa/\lambda$ in the limit of $\xi \ll \kappa, \lambda$. The action for the reference point $\kappa/\lambda =3$ has the analytic formula in Eq.~\eqref{eq:S4simple_3}.} 
    \label{fig:S4simple}
\end{figure}

For $\kappa = 3 \lambda$, the bounce solution has the fixed relation $\phi + S = v$. For this special parameter relation, $S_1$ can be calculated from a simple integration by noting that $dS/d\phi = - 1$ and $(\phi^*, S^*)$ is approximately the true minimum $(\pm v, 0)$. In the limit of $\xi \rightarrow 0$, 
\beqa
& S_1 (\kappa/\lambda = 3) = \dfrac{\sqrt{2}}{3}\,\sqrt{\lambda}\,v^3 ~,
\label{eq:S1simple_3} 
\\ \vspace{0mm}
& S_4(\kappa/\lambda = 3) = \dfrac{128\,\pi^2\,\lambda^2}{3\,\xi^3} ~. 
\label{eq:S4simple_3}
\eeqa
For different ratios of $\kappa/\lambda$, we use the program $\tt{FindBounce}$~\cite{Guada:2018jek} to numerically obtain the bounce profiles and report the obtained action $S_4(\kappa/\lambda)$ over the reference point $S_4(3)$ from \eqref{eq:S4simple_3} in Fig.~\ref{fig:S4simple}. 
Because $S_4$ generally changes only by $\mathcal{O}(1)$ for different choices of $\kappa/\lambda$, all later calculations will take $\kappa/\lambda=3$ as a fiducial value.

The tunneling rate (per four-dimensional volume) is~\cite{Coleman:1977py,Callan:1977pt} 
\beqa
\label{eq:tunneling-rate}
\gamma\,=\, \eta\, (2\,\lambda\,v^2)^2\,\left(\frac{S_4}{2\pi}\right)^2 \, e^{-S_4} ~, 
\eeqa
with $\eta$ as an order-one number that can be determined  from the functional determinant of fluctuations around the bounce; its precise value has negligible effects for later parts of this paper. The scale $2\,\lambda\,v^2$ is the mass square of the quantum field around the minima and sets the general scale for the tunneling rate. For the current minimal model, the tunneling rate $\gamma$ is time-independent. For later nonminimal models, $\gamma$ could be a function of time.

To estimate the percolation time, the fraction of space in the false vacuum as a function of time is
\beqa
\label{eq:false_vac_frac}
f(t) &=&\exp\left[-\frac{4\pi}{3}\int^t_{0}dt^\prime\,\gamma(t^\prime)\,a^3(t^\prime)\,r^3(t^\prime,t)\right] \\
&=&\exp\left[-\frac{3\pi\,\gamma\, t^4}{55}\right]\,,
\eeqa
where the comoving radius is
\begin{align}
r(t^\prime,t)=\int^t_{t^\prime}dt^{\prime\prime}\,\frac{1}{a(t^{\prime\prime})}=\frac{3t^{2/3}_\ast}{a_\ast}[t^{1/3}-(t^{\prime})^{1/3}] \, ,
\end{align}
assuming matter domination (MD) with the scale factor of $a=a_\ast(t/t_\ast)^{2/3}$ and the Hubble parameter is $H=2/(3t)$, which is valid for a wide range of time after matter-radiation equality. 
The percolation time $t_p$ is defined with $f(t_p)=e^{-0.34}$~\cite{PercolationThresh} and is 
\beqa
\label{eq:tpercolation}
t_p=\left(\frac{19}{3\pi\uu\gamma}\right)^{1/4} ~.
\eeqa

For the percolation time to be after the CMB recombination time $t_{\rm cmb} \approx 3.8 \times 10^5\,\mbox{yr}$ and before the current age of the universe $t_0 \approx 1.38 \times 10^{10}\,\mbox{yr}$, the tunneling rate satisfies 
\beqa
1.8\times10^{-33}\text{ eV}\lesssim \gamma^{1/4}\lesssim 6.8\times 10^{-29}\text{ eV} ~.
\eeqa
Choosing the lower end of $\gamma^{1/4} =1.8\times 10^{-33}$~eV for the current universe age and requiring $\Delta V = (2\times 10^{-3}\,\mbox{eV})^4$ to match the order of magnitude of the current CC, one has 
\beqa
\lambda = 6.6\times \left( \frac{2\times 10^{-3}\,\mbox{eV}}{v} \right)^6 \,, \qquad 
\xi = 4 \times \left( \frac{2\times 10^{-3}\,\mbox{eV}}{v} \right)^4 ~.
\eeqa
This suggests that the simple model in \eqref{eq:Z2-symmetric-potential} prefers to have the VEV scale $v$ around the CC scale for $\lambda \gtrsim \xi = O(1)$ and below the perturbative limit of $(4\pi)^2$.  

The number density of nucleated bubbles of true vacuum as a function of time is 
\beqa
n(t)=\frac{1}{a^3(t)}\int^t_{0} dt^\prime \gamma(t^\prime)\,f(t^\prime)\,a^3(t^\prime) =\left(\frac{55}{3\pi}\right)^{3/4}\frac{\gamma^{1/4}\left[\Gamma(\frac{3}{4})-\Gamma(\frac{3}{4},\frac{3\pi}{55}\gamma\uu t^4)\right]}{4t^2} ~.
\eeqa
The number of bubbles per Hubble patch at the percolation time is 
\beqa
N(t_p)&=&\frac{n(t_p)}{H^3(t_p)}=\left(\frac{1485}{\pi}\right)^{3/4}\frac{t_p\gamma^{1/4}\left[\Gamma(\frac{3}{4})-\Gamma(\frac{3}{4},\frac{3\pi}{55}\gamma\uu t^4_p)\right]}{32} \nonumber \\
&=&\frac{495\left[\Gamma(\frac{3}{4})-\Gamma(\frac{3}{4},1)\right]}{32\pi}\approx 1.9 ~,\label{eq:Ntn_vacuum}
\eeqa
which is a small number due to the slow bubble-nucleation rate. 

For the case with an even later quantum decay during a CC-dominated universe which may occur during a ``supercooled'' QT, bubble nucleation during both the matter-dominated and CC-dominated universe must be included. The Hubble parameter is approximated by
\beqa
H(t) \approx \begin{cases}
H_S\,\left(\dfrac{a(t)}{a_{\rm MV}}\right)^{-3/2}\,, & t < t_{\rm MV} \,,\\
H_S\,, & t > t_{\rm MV} \,.
\end{cases}
\eeqa
Here, $t_{\rm MV}$ is the time of matter-vacuum energy equality, and $H_S \equiv \sqrt{8\pi\,G\,V_S^0/3}$ with $G$ as the Newton constant and $V_S^0$ from \eqref{eq:V_S^0}. To use this Hubble parameter, we anticipate a late percolation time with $t_p > t_{\rm MV}$. Solving the Hubble equation, the scale factor is
\beqa
a(t) = \begin{cases}
a_{\rm MV}\left(\dfrac{t}{t_{\rm MV}}\right)^{2/3}\,, & t < t_{\rm MV} \,, \vspace{3mm} \\ 
a_{\rm MV}\,e^{H_S(t - t_{\rm MV})}\,, & t > t_{\rm MV} \,.
\end{cases}
\eeqa

The exponential argument for the fraction $f(t)$ has the following approximate formula in the limit of $t \gg t_{\rm MV}$:
\beqa
\ln[f(t)] &=& \frac{2\pi}{9}\,(11 - 18\,e^{-H_S\,t} + 9\,e^{-2\,H_S\,t} - 2\,e^{-3\,H_S\,t} - 6\,H_S\,t)\,\frac{\gamma}{H_S^4} ~, \nonumber \\
& \overset{H_S\,t \ll 1}{=}&  -\frac{\pi}{3}\,\gamma\,t^4 ~.
\eeqa
The percolation time with $f(t) \approx e^{-0.34}$ has the approximate solution 
\beqa
t_p \approx 
\begin{cases}
\left(\dfrac{1}{\pi\,\gamma}\right)^{1/4}, & \gamma^{1/4} \gg H_S \,,\\[10pt]
\dfrac{H_S^3}{4\pi\,\gamma}& \gamma^{1/4} \ll H_S ~. 
\end{cases}
\eeqa
Note that one needs to have $t_p > t_{\rm MV}$ to use the above formula.

During the supercooling or inflation period, the phase transition may not finish (see Refs.~\cite{Guth:1982pn,Turner:1992tz,Ellis:2018mja}). An approximate condition for the completion of the phase transition requires that the physical volume of the false vacuum $V_{\rm false}(t) = a(t)^3\,f(t)$ at around the percolation time decreases: 
\beqa
\frac{1}{V_{\rm false}}\,\dfrac{V_{\rm false}(t)}{dt}\Big|_{t=t_p} < 0 &\Rightarrow& 9\,H_S^4 - 4\,\pi\, (1-e^{-H_S\,t})^3 \,\gamma \,\big|_{t=t_p} < 0 \nonumber \\
 &\Rightarrow& \gamma^{1/4} \gtrsim  2.1 \, H_S ~. 
\label{eq:volume-constraint}
\eeqa

The number of bubbles per Hubble patch at the percolation time is 
\beqa
\label{eq:Ntn_vacuum_supercool}
N(t_p)=\frac{n(t_p)}{H_S^3} \approx  
0.7\,\times \, (\gamma^{1/4}/H_S)^{3} \qquad \quad \mbox{for}\quad \gamma^{1/4} \gtrsim 2.1\,H_S  \,.
\eeqa

During the matter-dominated universe, one has $H(t) = 2/(3\,t)$ or $H_S \approx 2/(3\,t_{\rm MV})$. For the quantum tunneling percolation time after $t_{\rm MV}$ or $t_p \gtrsim t_{\rm MV}$, one has
\beqa
t_p > t_{\rm MV} \quad \Rightarrow \quad  \gamma^{1/4}/H_S  \lesssim 2 ~. 
\label{eq:tp-over-tMV}
\eeqa
Comparing the two conditions in \eqref{eq:volume-constraint} and \eqref{eq:tp-over-tMV}, it appears that a vacuum transition cannot complete during a supercooled period, though more refined estimates for these equations could reveal a narrow window where this is possible.

Another consideration involves whether the transition has time to complete (\ie, have all of space in the true vacuum, which is a more stringent requirement than percolation) before the present day. An incomplete transition is estimated to occur when the average comoving distance between bubble nucleation sites is too large for the bubbles to meet before today: 
\begin{equation}
\label{eq:pt-completion}
    [n(t_p)]^{-1/3} (1+z_p) \gtrsim \Delta \tau = \int_0^{z_p} dz \, (1+z)^{-1} H_0^{-1} [\Omega_\Lambda+\Omega_m(1+z)^3]^{-1/2}
\end{equation}
where $\Delta \tau$ is the comoving distance from today to the redshift at which the transition occurs. Note that the dark energy density $\Omega_\Lambda$ is included here to obtain the correct expansion history between percolation and today. For example, with $\Omega_\Lambda=0.69$ and $\Omega_m=0.31$, $z_p \gtrsim 2.4$ for the transition to complete before today if it occurred during the MD era. Incomplete transitions are not necessarily problematic---if they reach percolation, they will eventually complete in the future---but the results presented in Sec.~\ref{sec:cosmology} will assume completed transitions for simplicity.

\subsection{Quantum tunneling + dark matter (QT+DM)}

Particles in different vacua can have different masses. One simple possibility is that a (subcomponent of) DM couples directly to the order-parameter field involved in the quantum tunneling process. In that case, we expect not only a change in the CC, but also corresponding changes in the DM energy density and the radiation energy density, in order to conserve the total energy. In this subsection, we introduce a simple DM variation of the QT model in the previous subsection. 

In this model, in addition to the potential terms in Eqs.~(\ref{eq:Z2-symmetric-potential}) and (\ref{eq:V-phiS-breaking}), the $S$ field (but not the $\phi$ field\footnote{Of course, some tuning is still present, as radiative corrections that couple $\chi, \bar{\chi}$ to $\phi$ must be suppressed. Or, such terms could be suppressed by symmetries of the theory: $\phi$ can still be real with a $\mathbb{Z}_2$ global symmetry, while $S$ and $\chi$ could be in some larger representation like $SU(2)$ more akin to how the Standard Model Higgs field couples to Standard Model fermions (the scalar potential would therefore be modified to replace $S^2$ with $|S|^2$). Indeed, a complex $S$ charged under the fermion number gauge symmetry is contemplated below.})
is assumed to have a Yukawa coupling to some Dirac fermion state $\chi( \bar{\chi})$ (making up only a percentage of all of DM), as
\begin{equation}
\label{eq:yukawa-dm}
    \mathcal{L} \supset y\, S \bar{\chi} \chi \, ,
\end{equation}
with $y$ the coupling constant.
This will result in this component of DM becoming massless after the transition to the $S=0$ vacuum [see \eqref{eq:tunneling}]. Therefore the $\chi$ component of DM will behave instead as DR after the transition to Vac$_\phi$. As we will see later, this may help in explaining the DESI+SN data.

The PT could proceed as a purely vacuum PT. In such a scenario, Eq.~(\ref{eq:Ntn_vacuum}) indicates that the number of bubbles nucleated per Hubble patch during the PT is of order one. 
As we will see in Section~\ref{sec:ISW}, late-time PTs are generally required to have many more bubble nucleation sites than this. Otherwise, they produce large ISW-like perturbations that are incompatible with observations of the CMB.

Therefore, we instead consider the contribution of finite density effects to the scalar potential. Let us assume that there is an initial $\chi \, \bar{\chi}$ particle-anti-particle asymmetry. If $\chi-\bar{\chi}$ annihilations have already frozen out, then the free energy density is $F = \Omega + \mu \, n$ with $\Omega$ the grand potential and $\mu$ the chemical potential, which is
\beqa
V_{\rm den}(n, S) = F(\mu, S) &=& g \int \frac{d^3p}{(2\pi)^3}\,E_p\, \Theta(\mu - E_p) ~, \nonumber \\
&=& \frac{g}{16\pi^2}\left[ \mu\,p_F (2p_F^2 + y^2\,S^2) - y^4\,S^4\ln\left(\frac{\mu + p_F}{|y\,S|}\right) \right] ~,
\eeqa
with $\mu = \sqrt{p_F^2 + y^2 S^2}$ and $g=2$ the degrees of freedom of $\chi$. The dark fermion number density is 
\beqa
n(\mu, S) = - \frac{\partial \Omega}{\partial \mu} = g \int \frac{d^3p}{(2\pi)^3}\, \Theta(\mu - E_p) = \frac{g\,p_F^3}{6\pi^2}~.
\eeqa
Substituting $p_F$ by $n$ and $\mu = \sqrt{p_F^2 + y^2 S^2}$, the finite-density potential has the nonrelativistic limit of $V_{\rm den}(n, S) = |y\,S|\,n$ with $p_F \ll |y \, S|$. 

$V_\text{den}$ contributes a positive contribution to the $S$ potential, raising the potential energy difference $\Delta V$ between Vac$_S$ and Vac$_\phi$. The effect of $V_\text{den}$ decreases with time as $n$ decreases. As a result, the corresponding tunneling action $S_4 \propto \Delta V^{-3}$ increases with time. Therefore, this model prefers to tunnel at an earlier time, rather than a later time (the subject of this work). This situation is similar to the pure QT case: the tunneling process is relatively slow, with only a few bubbles present at the percolation time. 
Even if the transition can be arranged to finish in the late universe, it would still be incompatible with CMB observations, as we will show later in Sec.~\ref{sec:ISW}. 

To change the behavior of the dependence of the effective potential energy in $n$, we also add the following effective four-fermion interaction
\beqa
\Delta V_4 = \frac{g_d^2}{M^2_A(S)} \bar{\chi}\gamma^\mu \chi \bar{\chi}\gamma_\mu \chi \qquad \mbox{with} \qquad M^2_A(S) = M^2_{A0} + g_d^2 \,|S|^2 ~.
\eeqa
This requires us to promote the scalar field $S$ to a complex one that is charged under the dark fermion number gauge symmetry $[U(1)_\chi]$. A simple Lagrangian to UV-complete this interaction could be 
\beqa
\mathcal{L} \supset - \frac{1}{4}\,F_{A\,\mu\nu}F_A^{\mu\nu} + D_\mu S D^\mu S^\dagger + \frac{1}{2}M^2_{A0} A_\mu A^\mu + g_d \, A_\mu \bar{\chi} \gamma^\mu \chi ~,
\eeqa
with $D_\mu S = \partial_\mu S + i g_d A_\mu S$, and $M^2_{A0}$ from other heavier scalar VEV contributions. In the limit of $g_d^2 |S|^2 \ll M^2_{A0}$ and taking $\bar{\chi} \gamma^0 \chi = n$, the corrected potential becomes
\beqa
\Delta V_4 = - \frac{g_d^4}{M^4_{A0}}\,|S|^2\, n^2 ~. 
\eeqa
This provides a negative contribution to the potential at larger density.

We first ignore the effects from the Yukawa coupling in Eq.~(\ref{eq:yukawa-dm}) on the total effective potential. We will check the self-consistency of this assumption later. As in the QT model, $V_S^0 > V_\phi$ such that at a very late time with $n\rightarrow 0$, the vacuum Vac$_S$ is metastable and may tunnel to the vacuum Vac$_\phi$. Including $\Delta V_4$, for a sufficiently large dark fermion number density $n$, the relative effective potential values of the two vacua $\mbox{Vac}_S$ and $\mbox{Vac}_\phi$ can be reversed such that $\mbox{Vac}_S$ can be the lower minimum. There then exists a critical density $n_c$ such that the two vacua are degenerate: 
\beqa
V_S^0 + \Delta V_4 = V_0 + \frac{\xi}{4 }\,v^4 -  \frac{g_d^4}{M^4_{A0}}\,|S|^2\, n_c^2 = V_0 \quad \Rightarrow \quad  n_c \approx \frac{\sqrt{\xi}}{2}\,\frac{v\,M_{A0}^2}{g_d^2} ~. 
\eeqa

When the number density is near to and slightly below the critical number density, the energy difference between the two vacua is
\beqa
\Delta V =V_S^0 + \Delta V_4 - V_\phi 
\approx \frac{g_d^4}{M^4_{A0}}\,(n_c^2 - n^2)\,v_s^2 ~,
\eeqa
where we have assumed that the finite-density effect induces a negligible change for the local VEV of $S$ around $v_s$. Using the four-dimensional Euclidean action for a vacuum PT in the thin wall approximation following Coleman \cite{Coleman:1977py,Callan:1977pt},
\beqa
S_4(n) = \frac{27\pi^2S^4_1}{2 \DV^3} \approx \frac{27 \pi^2\,S_1^4}{2}\, \left(\frac{M^4_{A0}}{g_d^4\,v^2 }\right)^3\,\frac{1}{(n_c^2 - n^2)^3} \equiv S_4(n=0)\,\frac{n_c^6}{(n_c^2 - n^2)^3} ~.
\eeqa
Here, $S_1$ can be calculated as in \eqref{eq:S1simple_3}, and $S_4(n=0)= 128\pi^2 \lambda^2/(3\xi^3)$ is the same as the QT case in \eqref{eq:S4simple_3}. 

Using the tunneling rate in \eqref{eq:tunneling-rate}, the false vacuum fraction is, similar to Eq.~(\ref{eq:false_vac_frac}),
\beqa
f(t) = \mbox{exp} \left[ - \frac{4\pi}{3}\,\int^t_{t_c}\,dt'\gamma(t')\,(t - t')^3 \right] ~,
\eeqa
where scale factor evolution is negligible for a faster tunneling process. Here, $t_c$ is the starting time of the tunneling process with $n=n_c$. The parameter $\beta/H\big|_{t_p}$ is used to quantify the nucleation speed, defined as
\beqa
\beta \equiv \frac{d}{dt}\ln{\gamma}\Big|_{t_p} \simeq - \frac{dS_4}{dt}\Big|_{t_p} ~.
\eeqa
Writing $\gamma(t') = \exp [\log \gamma(t')]$ and expanding $\log \gamma(t') = \log \gamma(t_p)+ (t'-t_p) \beta + \mathcal{O}[(t'-t_p)^2]$, the percolation time can be approximated by
\begin{equation}
    0.34 \approx \frac{4\pi}{3}\,\int^{t_p}_{t_c}\,dt' \gamma(t_p) \, e^{(t'-t_p) \beta} \, (t_p - t')^3 \approx 8 \pi \gamma(t_p) \beta^{-4} \, .
\end{equation}

Dark fermion number conservation within a comoving volume implies $n \propto a^{-3}$, giving $d\ln{n}/dt = - 3 H$. Therefore, 
\beqa
\frac{\beta}{H_p} \simeq 3 \frac{dS_4}{d\ln{n}}\Big|_{t_p} = 6 \frac{dS_4}{d\ln{n^2}}\Big|_{t_p} = 18\times\frac{27 \pi^2\,S_1^4}{2}\, \left(\frac{M^4_{A0}}{g_d^4\,v^2 }\right)^3\,\frac{n_p^2}{(n_c^2 - n^2_p)^4} = 18\, S_4 \frac{n_p^2}{n_c^2 - n^2_p}  ~.
\eeqa
with $H_p \equiv H\big|_{t_p}$.
Since $n_c - n_p \ll n_c$, one anticipates a large value for $\beta/H$ (even compared to the finite-$T$ phase transition). Letting $\epsilon \equiv (n_c^2-n_p^2)/n_c^2$, the solution for $\epsilon$ or $n_p$ is 
\begin{equation}
\label{eq:beta_finite_density_epsilon}
\epsilon \approx [S_4(n=0)]^{1/3} \left[ \log \left( \frac{\eta\,\lambda^2\,v^4 \epsilon^{10}}{0.34\times 13122 \, \pi S^2_4(0)\,H_p^4} \right) \right]^{-1/3} ~. 
 \end{equation}
For example, choosing $\eta=1$, $\lambda=6$, $\xi = 6$, $g_d=1$, $v=2\times 10^{-3}$~eV, $H_p\approx \sqrt{8\pi G (2 \times 10^{-3}\,\mbox{eV})^4/3}$, $M_{A0}=10^{-3}$~eV, $m_\chi = 5\times 10^{-5}$~eV, and $\fa = 0.008$ as the fraction of $\chi$ in DM, one has $\epsilon \approx 0.6$, $S_4 \approx 260$, $\beta/H_p \approx 7 \times 10^3$, and $\Delta V_4\approx (2.5\times 10^{-3})^4$ that is comparable to both $V_0$ and $V_S^0$.  One can see that the finite-density triggered quantum tunneling can be very quick with a large $\beta/H_p$. Returning to the assumption of neglecting $V_{\rm den}$ in the scalar potential, the contribution of the dark fermion mass to the effective potential is of order $\mathcal{O}(m_\chi^2\,n^{2/3})$, and is suppressed near $n_c$ for this benchmark point.

The nucleation density is similarly calculated as
\begin{equation}
\label{eq:nnuc}
    n_\text{nuc} = \int^{\infty}_{t_c}\,dt' \gamma(t') f(t') \approx (8 \pi)^{-1} \beta^3 \, ,
\end{equation}
which means that there are a large number of bubbles per Hubble patch at the percolation time with $N_{\rm nuc} \approx (\beta/H_p)^3/(8\pi)$. This significantly relaxes the transition completion bound in (\ref{eq:pt-completion}). For example, using $\Omega_\Lambda=0.69$ and $\Omega_m=0.31$ as before with $\beta/H_p=100$ (500), completion requires $z_p \gtrsim 0.03$ (0.006).

Two bounds should be addressed regarding the small fermion mass showing up in the calculation.
The first is the Tremaine-Gunn limit~\cite{Tremaine:1979we}, derived from the requirement that the DM number density obtained from the Local Group dwarf galaxy observations be no larger than the value that can be supported by the DM phase space.
The limit generically requires the fermionic DM mass to be heavier than several hundred eV~\cite{Hayashi:2016kcy}. Nevertheless, for the parameter range of interest, the light fermions should have already been turned into DR before the formation of these structures. Additionally, the limit is not effective on small subcomponents of DM.
The Tremaine-Gunn limit is thus not relevant.

The other constraint is the free-streaming constraint.
Too light a DM is likely to be relativistic in the early universe such that they can erase the density fluctuations and suppress the structure formation.
This is most commonly applied on sterile neutrinos, requiring their masses to be greater than around keV~\cite{Irsic:2017ixq}.
However, if the production mechanism can have the DM to be cold, this constraint can also be evaded. 
We will simply follow this assumption in later analysis, without specifying the underlying mechanism. This constraint is also weakened for DM subcomponents.

\subsection{Quantum tunneling + dark matter + domain wall (QT+DM+DW)}

Now let us add in the possibility of DW formation, keeping the coupling to DM the same as in the previous subsection. Here, the $\phi$ field is assumed to be real, so that DWs form during the PT due to the spontaneous breaking of the $\mathbb{Z}_2^\phi$ symmetry.

The number of DWs per Hubble patch produced is directly related to the number of bubbles nucleated during a PT, following the Kibble-Zurek mechanism \cite{Kibble:1976sj,Zurek:1985qw}. For a purely vacuum PT, this is
\begin{equation}
\label{eq:N_DW_vacuum}
    N_{\rm DW} \sim N(t_p) \sim \mathcal{O}(1) \, ,
\end{equation}
where Eqs.~(\ref{eq:Ntn_vacuum}) or (\ref{eq:Ntn_vacuum_supercool}) are used for the final equality. 
On the other hand, for a temperature- or density-dependent PT rate like in the previous subsection, the number of DWs per Hubble patch is
\beqa
N_{\rm DW} \sim N_{\rm nuc} \approx \frac{1}{8\pi}\, \left(\frac{\beta}{H(T_p)} \right)^3 \gg 1 ~,
\label{eq:N_DW}
\eeqa
where the results below Eq.~(\ref{eq:beta_finite_density_epsilon}) indicate that the final inequality is generically true.

The DWs have tension $\sigma \approx S_1$ in the thin-wall approximation. The DW energy density is estimated to be 
\beqa
\rho_{\rm DW} \sim N_{\rm DW}\, \sigma \, \frac{1}{\beta^2}\,H_p^3 \approx S_1\,\beta \quad  \Rightarrow \quad \frac{\rho_{\rm DW}}{\rho_\Lambda} \sim \frac{\sqrt{\lambda}\,v^3}{M_{\rm pl}\,\rho_\Lambda^{1/2}}\,\left(\frac{\beta}{H_p}\right) \sim \frac{v}{M_{\rm pl}}\, \left(\frac{\beta}{H_p}\right)  ~,
\eeqa
where the relation of $\rho_\Lambda \sim \lambda \,v^4$ is used for the last step.
One can see that to have $\rho_{\rm DW}/\rho_\Lambda\sim \mathcal{O}(1)$, we need to have a large value of the symmetry-breaking scale $v \sim M_{\rm pl}\,(H_p/\beta)$ and a tiny $\lambda$ to have the vacuum energy comparable to the CC or $\rho_\Lambda \sim \lambda \,v^4$.


\section{The cosmological implications of quantum tunnelings}
\label{sec:cosmology}

Possible cosmological constraints on the QT-involved models are discussed in this section.
At the background level, a QT changes the cosmic energy budget and thus influences the evolution of the various cosmological distances (the luminosity distance, the angular distance, etc.), which can be constrained by observations like BAO and SN distances.
Meanwhile at the perturbation level, the fluctuations sourced by the true vacuum bubbles nucleated during the tunneling will also leave imprints on the CMB anisotropy spectrum.
We thus examine whether the models presented in Sec.~\ref{sec:model} are compatible with the corresponding observational data, and present the preferred and ruled-out model parameter spaces.
We first discuss how the relevant cosmological distances are influenced by the QT in Sec.~\ref{sec:background_evo}, and then check the compatibility of the models with the DESI DR2 anomaly (DESI+SN+CMB) in Sec.~\ref{sec:DESI_fitting}, with the posterior distributions of the models presented.
Lastly, in Sec.~\ref{sec:ISW} we check the constraints on the QT time and energy budget from the CMB anisotropy measurements.

\subsection{Cosmological evolution and distances with a QT}
\label{sec:background_evo}

The cosmic evolution of the tunneling is parametrized as follows. Ignoring the duration of the tunneling, we take the vacuum transition to occur instantaneously at redshift $\zt$, corresponding to the scale factor $\at=1/(1+\zt)$.
As modeled in the previous section, the tunneling can transform a fraction of cold dark matter (CDM) and vacuum energy into DWs plus DR.
The remaining pieces of the cosmic energy budget: baryons, Standard Model radiation, neutrinos, and the cosmological constant $\Lambda$ (\ie, the vacuum energy after the QT, denoted as $V_0$ in the previous section), remain unchanged before and after the transition. The cosmic evolution of the Hubble parameter $H$ can then be written in terms of the components of the cosmic energy budget as
\begin{align}
H^2=H^2_0\times\begin{cases}
\left(\Omega_{b,0}+\Omega_c\right)a^{-3}+\Omega_{r,0\uu }a^{-4}+\Omega_\nu(a)+\Omega_{\Lambda,0}+\Omega_V & a<a_t\\
(\Omega_{b,0}+\Omega_{c,0})a^{-3}+(\Omega_{r,0}+\Omega_{dr,0})a^{-4}+\Omega_{dw,0}\uu a^{-1}+\Omega_\nu(a)+\Omega_{\Lambda,0} & a>a_t
\end{cases}\,,\label{eq:Hsq_all}
\end{align}
where $\Omega_{i,0}=\rho_{i,0}/\rho_{{\rm crit},0}$ is the fractional energy density today of component $i=b$, $c$, $r$, $dr$, $dw$, and $\Lambda$ corresponding to baryon, CDM, radiation, DR, DW, and the CC, respectively. $\Omega_\nu(a)=\rho_\nu(a)/\rho_{{\rm crit},0}$ is the ratio between the cosmic neutrino energy density and the {\it present time} cosmic energy density. Similarly, $\Omega_V=\Delta V / \rho_{{\rm crit},0}$, with $\Delta V$ the difference in dark energy between the two vacua of the QT.
The DWs are assumed to be frustrated, with energy density redshifting like $a^{-1}$. This is because they are created in the late universe, and may not have sufficient time to reach the scaling regime.
Note that we use an unconventional normalization scheme where all the $\Omega(a)$s are normalized with respect to $\rho_{{\rm crit},0}$ rather than the cosmic energy density $\rho_{{\rm crit}}(a)$ at the scale factor $a$.
With such a choice, it is possible that $\Omega_V$ could be greater than one due to the dilution of $\rho_c$ over the cosmic redshift. Similarly, the flat universe constraint
\begin{align}\label{eq:budget_normalization}
\Omega_{b,0}+\Omega_{c,0}+\Omega_{r,0}+\Omega_{dr,0}+\Omega_{dw,0}+\Omega_{\nu}(a=0)+\Omega_{\Lambda,0}=1\,,
\end{align}
is imposed on the energy fractions $\Omega_{i,0}$, but not necessarily at an earlier cosmic time for the energy fractions $\Omega_i(a)$.

Assuming $\chi$, the dark matter converted to DR by the QT, makes up a fraction $\fa$ of all CDM before the tunneling, the CDM abundances before and after the transition are related by
\begin{align}\label{eq:Omega_c_cont}
\Omega_c=\frac{1}{1-\fa}\Omega_{c,0}\,.
\end{align}
If a fraction $\xdw$ of the transformed energy (\ie, those in $\chi$ and $\Delta V$) is stored as DWs with the remaining energy stored in DR, then energy conservation at the tunneling implies
\begin{align}\label{eq:Omega_dw_cont}
&\frac{\Omega_{dw,0}}{a_t}=\xdw\left(\frac{\fa}{1-\fa}\frac{\Omega_{c,0}}{a^3_t}+\Omega_V\right)\,,&&\frac{\Omega_{dr,0}}{a^4_t}=(1-\xdw)\left(\frac{\fa}{1-\fa}\frac{\Omega_{c,0}}{a^3_t}+\Omega_V\right)\,.
\end{align}
The full model is thus left with 7 different degrees of freedom: $(H_0,\,\Omega_{b,0},\,\Omega_{c,0},\,\zt,\,\fa,\,\xdw,\,\Omega_V)$.

A direct result of a late-time QT is the modification of the cosmic distances relevant for BAO and SN light curve measurements.
For BAO, the corresponding constraints could be imposed on the comoving distance
\begin{align}\label{eq:DM}
    D_M(z)=\int^z_0\dfrac{dz^\prime}{H(z^\prime)}\,,
\end{align}
the Hubble distance
\begin{align}\label{eq:DH}
D_H(z)=\dfrac{1}{H(z)}\,,
\end{align}
and their combination $D_V=(D^2_MD_H)^{1/3}$~\cite{DESI:2025zgx}.
SN, on the other hand, directly constrain the luminosity distance $D_L(z)=(1+z)D_M(z)$ via the distance modulus
\begin{align}
\mu=5\log_{10}\left(\frac{D_L}{10\text{ pc}}\right)\,.
\end{align}

\begin{figure}[t!]
    \centering
    \includegraphics[width=0.99\linewidth]{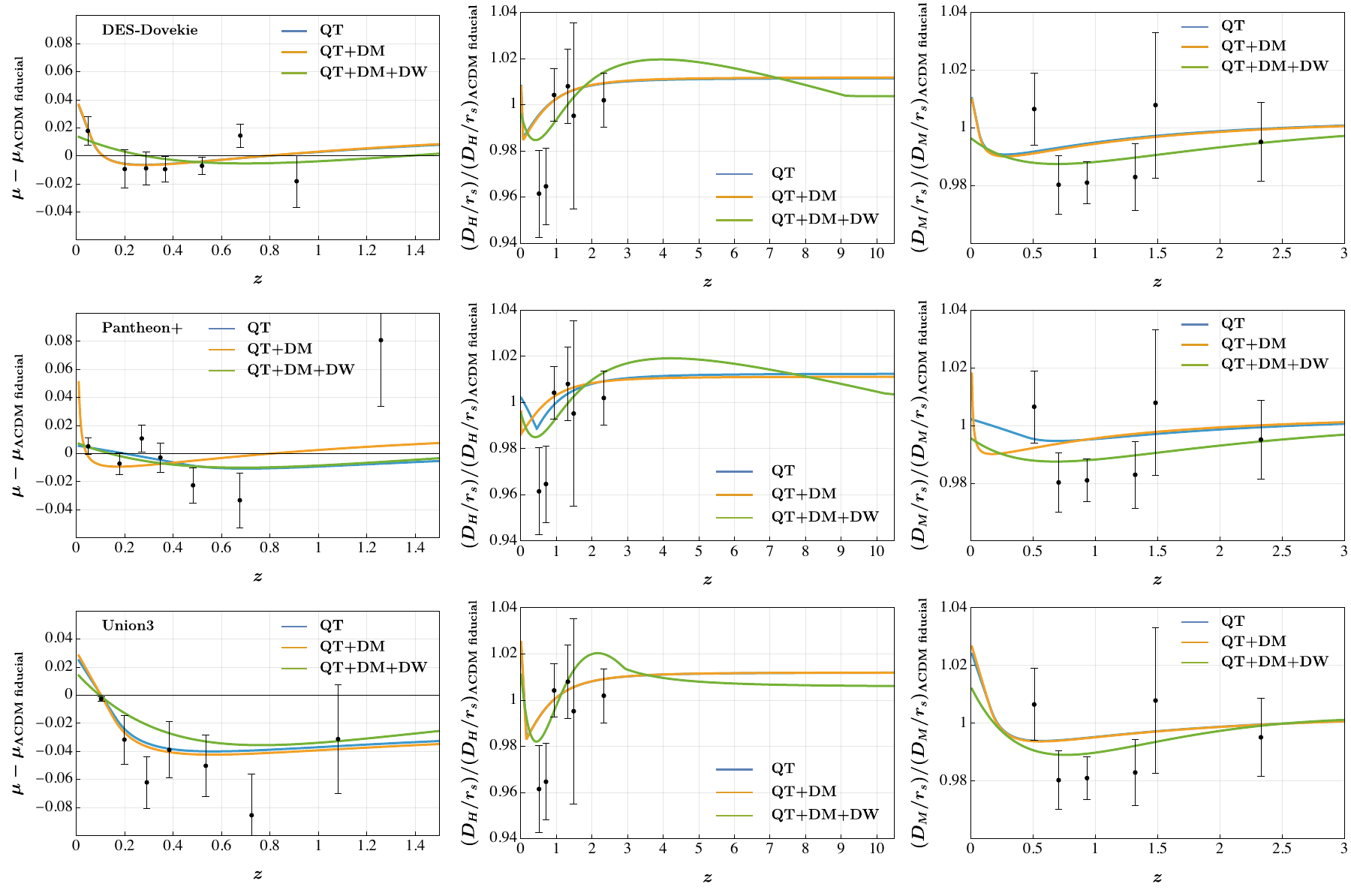}
    \caption{Hubble diagrams showing the comparison of $\mu$, $D_H/r_s$, and $D_M/r_s$ between the QT-involved models and the $\Lambda$CDM model, in terms of the cosmic redshift $z$. 
    The comparison is made between the fiducial $\Lambda$CDM model from the Planck 2018 results and the best-fit points of the QT models (second-best for QT+DM for visualization purpose).
    Different rows correspond to different SN datasets chosen for the joint fitting, as labeled in the first plot of each row.
    The blue, orange, and green curves in each panel represent the results of the QT, QT+DM, and QT+DM+DW models, respectively.
    Note that some blue curves are largely hidden by the orange curves, due the similarity between the QT and QT+DM model fitting results.
    The experiment data (SN for the left column and DESI DR2 for the middle and right columns) are shown in black.
    The error bars of the data are chosen to be the square root of the covariance matrix. 
    Model parameters of the curves shown in the figure as well as the rebinning method of the SN data are given in Appendix~\ref{app:binning}.
    }
    \label{fig:hubble_diag}
\end{figure}

Hubble diagrams of some of these quantities comparing various models and experimental data are shown in Fig.~\ref{fig:hubble_diag}.
The fiducial $\Lambda$CDM model uses the Planck 2018 TT, TE, EE+lowE+lensing best fit parameters~\cite{Planck:2018vyg} as in the DESI DR2 analysis~\cite{DESI:2025zpo}, while our own best-fit parameters are chosen for the other models.
Different rows of the panels correspond to different SN datasets.
The left and middle columns show the evolution of $D_H/r_s$ and $D_M/r_s$ in the QT-involved models, normalized against the corresponding values in the fiducial $\Lambda$CDM model,  with $r_s$ the comoving sound horizon when recombination happens.
The black points in these panels are the DESI DR2 BAO measurements~\cite{DESI:2025zgx}, whose error bars are chosen to be the square root of the corresponding diagonal entries in the covariance matrix.
The right column shows the difference of the SN distance modulus between the QT models or the data and the fiducial model.
Due to the sizes of the SN datasets, for visualization we reorganize the datasets into the same seven redshift bins.
The model curves and the data points of the same SN dataset are shifted by a constant offset such that they share the same weighted mean.
Details of the SN binning are given in Appendix~\ref{app:binning}.
The fitting and data analysis, on the other hand, are performed with the full dataset and will be elaborated in the following subsection.

By eye, it is reasonably clear from Fig.~\ref{fig:hubble_diag} that the $\Lambda$CDM model does not provide a good fit to the data because many data points differ from unity in the first two columns and zero in the last column. The QT+DM+DW model appears to provide the best fit, though this is expected since it has the most free parameters. The following subsection will quantify how well each of these models explain the data, taking into account the number of free parameters in each.

\subsection{Compatibility with the DESI DR2 anomaly}
\label{sec:DESI_fitting}

The data analysis is performed in the following way. 
The full analysis pipeline is implemented within the cosmological Bayesian analysis framework \texttt{Cobaya}~\cite{Torrado:2020dgo,2019ascl.soft10019T}, which provides interfaces to both the posterior sampler (including the datasets and the likelihood functions) as well as the cosmological theory tools.
For the theory tools, we use \texttt{CLASS}~\cite{Blas:2011rf} to track the cosmic evolution, where the ``background'' module is modified to accommodate the additional components of the energy budget and the tunneling.
A flat universe is assumed throughout the analysis.
The radiation energy density is kept the same as in the $\Lambda$CDM scenario to be consistent with the CMB temperature measurement.
Two massless and one massive neutrino species with mass 0.06 eV are assumed, and $N_{\rm eff}=3.044$. 

The posterior of the inference comes from the combination of the DESI DR2 BAO data, the Type Ia SN distance measurements, as well as the CMB measurements.
DESI DR2 constrains the models through the measurements of $D_M(z)$, $D_H(z)$, and $D_V(z)$.
SN distance measurements constrain the modification to $D_L(z)$ via the measurement of the SN distance modulus $\mu$.
Three different SN distance measurement datasets are considered in our analysis: Pantheon+~\cite{Scolnic:2021amr,Brout:2022vxf}, Union3~\cite{Rubin:2023jdq}, and the reanalyzed DES SN light curve dataset DES-Dovekie~\cite{DES:2025sig}, each of which is combined independently with the BAO and CMB measurements to obtain the likelihood. 

For the CMB, we consider a compressed version of the whole analysis for simplicity.
With the inference parameters $H_0$, $\Omega_{b,0}$, and $\Omega_{c,0}$ replaced by the angular size of the comoving sound horizon $\theta_s$, the abundance of the baryon matter $\omega_b=\Omega_{b,0}h^2$, and that of CDM $\omega_c=\Omega_{c}h^2$, the full CMB analysis has been shown to be well approximated by a multivariate Gaussian prior that marginalizes over all the other fitted parameters~\cite{DESI:2025zgx}.
The mean and covariance of the prior are~\cite{Braglia:2025gdo}
\begin{align}\begin{aligned}
&\mathbf{\mu}(100\,\theta_s,\,\omega_b,\,\omega_c)=(1.04103,\,0.02223,\,0.1192)\,,\\[2mm]
&\mathbf{\Sigma}=10^{-8}\times\begin{pmatrix}
6.62099420 & 1.24442058 & -13.1731741\\
1.24442058 & 2.13441666 & -11.5345007\\
-13.1731741 & -11.5345007 & 169.7763
\end{pmatrix}\,.\label{eq:mu_and_cov}
\end{aligned}\end{align}
Despite being different than the combination of parameters $(\theta_\ast,\,\omega_{bc},\,\omega_b)$ used in the DESI DR2 analysis~\cite{DESI:2025zgx}, the parameters we adopted provide similar results~\cite{Braglia:2025gdo} for the same model.
The priors of the inference parameters are summarized in Table~\ref{tab:prior}. 
Note that this prior space is not entirely physical, as some parameter combinations can evaluate to negative energy density fractions $\Omega_i$ and $\Omega_{i,0}$ with Eqs.~(\ref{eq:budget_normalization}--\ref{eq:Omega_dw_cont}).
The posterior is set to 0 manually during the inference when this occurs.
Also, it is possible that the chosen $\theta_s$ cannot provide a valid Hubble parameter $H_0$ with given $\omega_b$, $\omega_c$, $\zt$, $\fa$, $\xdw$, and $\Omega_V$, also suggesting the sampled parameter combination to be unphysical.
This is resolved by the interface between \texttt{CLASS} and \texttt{Cobaya}. 

\begin{table}
\centering
\begin{tabular}{c|c}
\hline
parameter & prior \\
\hline
$100\,\theta_s$ & $\mc{U}[1.03,\,1.05]$\footnotemark\\
$\omega_b$ & $\mc{U}[0.021,\,0.024]$\\
$\omega_c$ & $\mc{U}[0.1,\,0.14]$\\
$\zt$ & $\mc{U}[0.01,\,20]$\\
$\fa$ & $\mc{U}[0, 0.5]$\\
$\xdw$ & $\mc{U}[0,\, 0.5]$\\
$\Omega_V$ & $\mc{U}[0,\,5]$\\
\hline
\end{tabular}
\caption{The priors of the inferred parameters, where $\mc{U}$ stands for the uniform prior.}
\label{tab:prior}
\end{table}
\footnotetext{The uniform priors of $100\,\theta_s$, $\omega_b$, and $\omega_c$ are not in conflict with the mean and covariance discussed around Eq.~\eqref{eq:mu_and_cov}, which is included as an \textit{additional} exterior prior in \texttt{Cobaya}. The uniform priors provide a common offset for all inferred posteriors and hence will not influence the Bayes factor.}

The sampling of the high dimensional parameter space is performed by using the nested sampler \texttt{polychord}~\cite{Handley:2015vkr,Handley:2015fda}. Specifically, we use the settings \texttt{nlive=75d} and \texttt{num\_repeat=5d} to ensure that the number of sampled inference points is sufficient to cover the full parameter space.
The best fit model parameters and their associated $\chi^2$ likelihood are determined using the \texttt{Py-BOBYQA} minimizer~\cite{Cartis:2018jxl} integrated into \texttt{Cobaya}.
The Bayesian evidence $\ln\mc{Z}$ of the corresponding model is automatically computed by \texttt{polychord}, with the volume of the prior space calculated by using the one-likelihood provided by \texttt{Cobaya}, enabling model comparison through the Bayes factor $\ln\mc{B}=\Delta\ln\mc{Z}$.
Preferences among the candidate models are evaluated according to the Jeffery's scale~\cite{kass1995bayes}.
For a more panoramic comparison of each model's performance, we also report the difference between the models' Akaike information criterion (AIC)~\cite{Liddle:2007fy} and deviance information criterion (DIC)~\cite{spiegelhalter2002bayesian} with respect to the $\Lambda$CDM model.
The AIC difference is defined as $\Delta\text{AIC}=\Delta\chi^2+2\Delta p$, where $\Delta p$ is the parameter number difference between the two models. 
The DIC difference, on the other hand, is defined as $\Delta\text{DIC}=2\,\Delta\overline{\chi^2(\theta)}-\Delta\chi^2(\overline{\theta})$, where the overline indicates the expectation value, and $\theta$ indicates the set of model parameters.
The two information criteria directly or indirectly take into account the model parameter number in the comparison, penalizing models with more degrees of freedom.

Results of the data analysis are reported in Tables~\ref{tab:inference_result_DES}, \ref{tab:inference_result_Pan}, and \ref{tab:inference_result_Uni}, summarized according to the SN dataset (DES-Dovekie, Pantheon+, and Union3, respectively).
The $\Lambda$CDM model and the CPL parametrization are also shown.
Note that although the CMB constraints are included as a Gaussian \textit{prior} (implemented as an external prior in \texttt{Cobaya}),
the purpose of this data compression is to incorporate the full CMB \textit{likelihood} in a convenient way. 
The relevant inference result is therefore also added to the total $\chi^2$, which is calculated as $\chi^2_{\rm total}=\chi^2_{\rm BAO}+\chi^2_{\rm SN}+\chi^2_{\rm CMB}$, with $\chi^2_{\rm CMB}=2(-\ln\pi_{\rm CMB})$ where $\pi_{\rm CMB}$ is the prior from the CMB.

\begin{table}[t]
\renewcommand{\arraystretch}{1.2}
\centering
{\footnotesize
\begin{tabular}{c|c|c|c|c|c}
\hline
& $\Lambda$CDM (3) & CPL (5) & QT (5) & QT+DM (6) & QT+DM+DW (7)\\
\hline
$100\,\theta_s[\times1000]$ &
$1041.25^{+0.23}_{-0.24}$ &
$1041.13\pm 0.24$ &
$1041.27^{+0.21}_{-0.24}$&
$1041.15\pm 0.25$ &
$1041.10\pm 0.26$ \\
$\omega_b[\times1000]$ & 
$22.39\pm 0.10$ & 
$22.29\pm 0.13$ & 
$22.40\pm 0.12$& 
$22.30^{+0.14}_{-0.15}$ & 
$22.27^{+0.15}_{-0.14}$ \\
$\omega_c[\times1000]$ & 
$117.10^{+0.60}_{-0.62}$ & 
$118.36^{+0.84}_{-0.96}$ & 
$116.96^{+0.60}_{-0.66}$ & 
$118.12^{+1.30}_{-1.23}$ & 
$118.53\pm 1.31$ \\
$\zt$ & - & - &
unconstrained &
unconstrained &
$8.87^{+5.26}_{-3.60}$ \\
$\Omega_V$ & - & - & unconstrained & unconstrained & unconstrained \\
$\fa$ & - & - & - &
unconstrained ($<0.072\,(3\sigma)$) &
$0.125^{+0.118}_{-0.056}$ \\
$\xdw$ & - & - & - & - & $0.066^{+0.062}_{-0.031}$ \\
\hline
$w_0$ & - & $-0.818^{+0.059}_{-0.053}$ & - & - & - \\
$w_a$ & - & $-0.641^{+0.215}_{-0.248}$ & - & - & - \\
\hline
$\Delta\chi^2_{\text{BAO}}$ & - & $-3.01$ & 
$-1.49$ & 
$-1.48$ & $-3.43$  \\
$\Delta\chi^2_{\text{SN}}$ & - & $-5.63$ & 
$-7.75$ & 
$-7.75$ & $-4.60$ \\
$\Delta\chi^2_{\text{CMB}}$ & - & $-2.60$ & 
$0.80$ & 
$0.72$ & $-3.60$ \\
$\Delta\chi^2_{\text{total}}$ & - & $-11.23$ & 
$-8.44$ & 
$-8.51$ & $-11.64$\\
\hline
$\Delta\text{AIC}$ & - & $-7.23$ & 
$-4.44$ & 
$-2.51$ & $-3.64$ \\
$\Delta\text{DIC}$ & - & $-7.15$ & 
0.45 & 
$-16.35$ & $-15.53$\\
$\ln\mc{B}$ & - & $-0.23$ & 
$-0.88$ & 
$-2.64$ & $0.77$\\
\hline
\end{tabular}
}
\caption{Results of the data analysis on DES-Dovekie+DESI BAO+compressed CMB for various models. The number of model parameters are given in the bracket after the model name in the first line of the table. The median and $1\sigma$ range of the model parameters are first reported. Parameters marked ``unconstrained'' have no constraints on their 1D posterior distributions, but may be constrained when considering 2D posteriors, see Figs.~\ref{fig:triangle_5param},~\ref{fig:triangle_6param}, and~\ref{fig:triangle_7param}. For example, for $\fa$ in the QT+DM model we provide in parentheses the one-sided $3\sigma$ constraint after manually excluding samples with $\zt<1$. The $\chi^2$ performances of the best-fit points of each model compared to $\lcdm$ are provided next, and then the model evaluation criteria.}
\label{tab:inference_result_DES}
\end{table}

\begin{table}[]
\renewcommand{\arraystretch}{1.2}
\centering
{\footnotesize
\begin{tabular}{c|c|c|c|c|c}
\hline
& $\Lambda$CDM (3) & CPL (5) & QT (5) & QT+DM (6) & QT+DM+DW (7)\\
\hline
$100\theta_s[\times1000]$ & 
$1041.26^{+0.23}_{-0.20}$ & 
$1041.15^{+0.24}_{-0.25}$ & 
$1041.27^{+0.21}_{-0.24}$ & 
$1041.13^{+0.23}_{-0.27}$ & 
$1041.10^{+0.25}_{-0.26}$ \\
$\omega_b[\times1000]$ & 
$22.40^{+0.11}_{-0.14}$ & 
$22.31\pm 0.13$ & 
$22.40\pm 0.12$& 
$22.28^{+0.13}_{-0.14}$ & 
$22.28^{+0.14}_{-0.15}$ \\
$\omega_c[\times1000]$ & 
$116.95^{+0.72}_{-0.63}$ & 
$118.08^{+0.98}_{-1.04}$ & 
$116.87^{+0.63}_{-0.69}$ & 
$118.42^{+1.29}_{-1.18}$ & 
$118.53^{+1.30}_{-1.24}$ \\
$\zt$ & - & - & 
unconstrained &
unconstrained &
$8.84^{+5.67}_{-3.79}$ \\
$\Omega_V$ & - & - & unconstrained & unconstrained & unconstrained \\
$\fa$ & - & - & - & 
unconstrained ($<0.079\,(3\sigma)$) & 
$0.121^{+0.131}_{-0.057}$ \\
$\xdw$ & - & - & - & - & $0.065^{+0.072}_{-0.033}$ \\
\hline
$w_0$ & - & $-0.850^{+0.058}_{-0.048}$ & - & - & - \\
$w_a$ & - & $-0.536^{+0.219}_{-0.199}$ & - & - & - \\
\hline
$\Delta\chi^2_{\text{BAO}}$ & - & $-2.59$ & 
$-2.34$ & 
$-0.29$ & $-3.11$ \\
$\Delta\chi^2_{\text{SN}}$ & - & $-2.90$ & 
$-3.60$ & 
$-3.67$ & $-3.38$ \\
$\Delta\chi^2_{\text{CMB}}$ & - & $-1.95$ & 
$1.51$ & 
$0.35$ & $-3.23$ \\
$\Delta\chi^2_{\text{total}}$ & - & $-7.44$ & 
$-4.43$ & 
$-3.61$ & $-9.73$ \\
\hline
$\Delta\text{AIC}$ & - & $-3.44$ & 
$-0.43$ & 
$2.39$ & $-1.88$ \\
$\Delta\text{DIC}$ & - & $-3.43$ & 
$0.07$& 
$0.18$ & $-14.71$ \\
$\ln\mc{B}$ & - & $-2.09$ & 
$-0.70$ & 
$-2.24$ & $0.48$ \\
\hline
\end{tabular}
}
\caption{Same as Table.~\ref{tab:inference_result_DES}, with the SN dataset being Pantheon+.}
\label{tab:inference_result_Pan}
\end{table}

\begin{table}[]
\renewcommand{\arraystretch}{1.2}
\centering
{\footnotesize
\begin{tabular}{c|c|c|c|c|c}
\hline
& $\Lambda$CDM (3) & CPL (5) & QT (5) & QT+DM (6) & QT+DM+DW (7)\\
\hline
$100\theta_s[\times1000]$ & 
$1041.27^{+0.22}_{-0.27}$ & 
$1041.10\pm 0.25$ & 
$1041.29^{+0.22}_{-0.24}$ & 
$1041.12\pm 0.25$ & 
$1041.08^{+0.26}_{-0.25}$ \\
$\omega_b[\times1000]$ & 
$22.38^{+0.12}_{-0.10}$ & 
$22.28^{+0.14}_{-0.15}$ & 
$22.40\pm 0.12$ & 
$22.29^{+0.15}_{-0.14}$ & 
$22.26^{+0.15}_{-0.14}$ \\
$\omega_c[\times1000]$ & 
$116.99^{+0.59}_{-0.66}$ & 
$118.54^{+0.97}_{-1.10}$ & 
$116.85^{+0.63}_{-0.65}$ & 
$118.27^{+1.27}_{-1.37}$ & 
$118.67^{+1.34}_{-1.29}$ \\
$\zt$ & - & - & 
unconstrained &
unconstrained &
$7.56^{+5.02}_{-2.78}$ \\
$\Omega_V$ & - & - & unconstrained & unconstrained & unconstrained \\
$\fa$ & - & - & - & 
unconstrained ($<0.085\,(3\sigma)$) & 
$0.190^{+0.149}_{-0.093}$ \\
$\xdw$ & - & - & - & - & $0.088^{+0.072}_{-0.043}$ \\
\hline
$w_0$ & - & $-0.680^{+0.090}_{-0.089}$ & - & - & - \\
$w_a$ & - & $-1.001^{+0.298}_{-0.351}$ & - & - & - \\
\hline
$\Delta\chi^2_{\text{BAO}}$ & - & $-4.07$ & 
$-1.85$ & 
$-1.85$ & $-3.18$ \\
$\Delta\chi^2_{\text{SN}}$ & - & $-6.94$ & 
$-8.52$ & 
$-8.52$ & $-6.43$ \\
$\Delta\chi^2_{\text{CMB}}$ & - & $-2.89$ & 
$1.05$ & 
$0.80$ & $-3.21$ \\
$\Delta\chi^2_{\text{total}}$ & - & $-13.90$ & 
$-9.31$ & 
$-9.57$ & $-12.82$ \\
\hline
$\Delta\text{AIC}$ & - & $-9.90$ & 
$-5.31$ & 
$-3.57$ & $-4.92$ \\
$\Delta\text{DIC}$ & - & $-9.61$ & 
0.22 & 
$-15.10$ & $-16.04$ \\
$\ln\mc{B}$ & - & $2.41$ & 
$-0.06$ & 
$-1.70$ & $2.27$ \\
\hline
\end{tabular}
}
\caption{Same as Table.~\ref{tab:inference_result_DES}, with the SN dataset being Union3.}
\label{tab:inference_result_Uni}
\end{table}

Complementary to the tables, in Figs.~\ref{fig:triangle_5param},~\ref{fig:triangle_6param}, and~\ref{fig:triangle_7param} we show the posterior distributions of the data for the three QT-involved models,
with the different choices of SN dataset overlaid.
In general, the QT and QT+DM models are not strongly supported by the data compared to the $\lcdm$ model or CPL parametrization. Some $\zt$-dependent bounds can be set on their parameters. On the other hand, the QT+DM+DW model improves upon the $\lcdm$ model and performs similarly to or better than the CPL parametrization. More details of the performance of each model are discussed below.

\subsubsection{QT}
\label{sec:QT_fitting}

\begin{figure}[t!]
    \centering
    \includegraphics[width=0.7\linewidth]{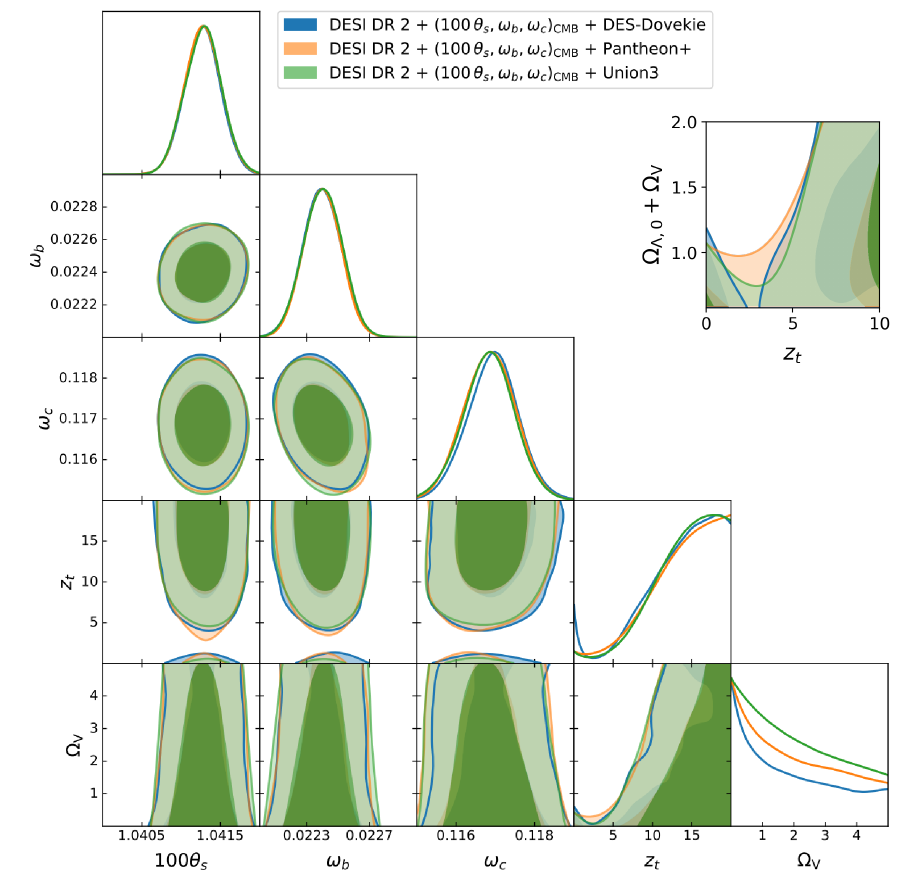}
    \caption{Posterior distribution for the parameters in the QT model for the likelihood combination of SN+DESI BAO+compressed CMB. The distributions for the three SN datasets are overlaid, where DES-Dovekie, Pantheon+, and Union3 are shown in blue, orange and green, respectively. 
    The total vacuum energy abundance today if the PT were not to complete, $\Omega_{\Lambda,0}+\Omega_V$, is dependent on the other parameters in the triangle plot; its posterior distribution with $z_t$ is shown as a separate panel (upper right). Because the PT does complete at $z_t>0$, it can be greater than unity without overclosure.
    }
    \label{fig:triangle_5param}
\end{figure}

As a special case of the full model with no DM mass change and no DW formation after the QT, this model contains 5 free parameters for fitting, with $\fa=\xdw=0$ imposed.
The posterior distribution of the model parameters is given in Fig.~\ref{fig:triangle_5param}.
Constraints on the model parameters and the model performances on different SN datasets are given in the middle column of Tables~\ref{tab:inference_result_DES},~\ref{tab:inference_result_Pan}, and~\ref{tab:inference_result_Uni}, respectively.

As seen in the posterior distributions, the two additional model parameters compared to the $\lcdm$ model, $\zt$ and $\Omega_V$, are not well constrained over the examined prior range.
$\Omega_V$ shows a largely flat distribution.
The tunneling redshift $\zt$, on the other hand, exhibits a bimodal feature for all three SN datasets, allowing either small or larger $z_t$ in the sampled parameter range.
Given that the DESI BAO measurements are generically made on redshift $0.5<z<2.5$, an intuitive interpretation is that the BAO measurements do not prefer for a QT to occur within their redshift range, and the tunneling should happen either rather early or rather late.
The small-$\zt$ region accommodates the best-fit point of the model.
At large $\zt$, the peak in the 1D posterior distribution at $\zt\sim17$ is a byproduct of the smoothing of the kernel density estimation and is therefore not physical.
In fact, we have checked that $\zt$ has a flat distribution on $\zt\gtrsim 20$ with an inference using $\mc{U}(0, 50)$ for the prior of $\zt$.

A remark on the joint distribution of  $\zt$ and $\Omega_V$ inferred from this analysis.
The two parameters are marked as ``unconstrained'' in Tables~\ref{tab:inference_result_DES},~\ref{tab:inference_result_Pan}, and~\ref{tab:inference_result_Uni} because their 1D posterior distributions are open or flat. Despite this, it is clear that a boundary exists in their joint distribution.
If their prior ranges can be narrowed down by some physical arguments, an upper or lower limit may be derived for $\Omega_V$ and $\zt$, respectively.
For example, if $\zt<10$, based on the results of Fig.~\ref{fig:triangle_5param}, $\Omega_V\lesssim 3$ can be derived at 2-$\sigma$ confidence level.
Such possibilities may be realized if new cosmological observations and evidence are included in the data analysis.
As another example, if $\zt <1$, examining the fractional change of vacuum energy (\ie, $\Omega_V/(\Omega_V+\Omega_{\Lambda,0})$), one may find that 95\%(68\%) of the samples have $\Omega_V/(\Omega_V+\Omega_{\Lambda,0})<0.52(0.22)$, with the median of the ratio at $\sim 0.17$.
In other words, a considerable change of vacuum energy as large as 50\% is still allowed by the BAO and SN distance measurements.
However, such small $z_t$ is severely constrained by CMB anisotropies (see Sec.~\ref{sec:ISW}) and by the requirement that the phase transition completes, as described around Eq.~\eqref{eq:pt-completion}.

In terms of the best-fit point (\ie, smallest $\chi^2$), this model provides a fit better than the basic $\lcdm$ model, but worse than the CPL parametrization and the full QT+DM+DW model. 
The model is further penalized in the $\Delta$AIC score for its extra parameters, meaning it is not substantially statistically supported.
The Bayes factor $\ln\mc{B}$ provides a similar conclusion, which actually slightly disfavors the QT model.
The $\Delta$DIC score also shows a slight disfavor for the QT model (note that a negative $\Delta$DIC indicates a preference, similar to $\Delta\chi^2$).
However, due to the bimodal nature of the posterior distribution for $z_t$, the DIC may not be a meaningful evaluation criterion\footnote{\label{fn:DIC-bimodality} This can be seen by realizing that the averaged parameter $\bar\theta$ is pulled by both regions of the $z_t$ posterior distribution, and therefore resides around the ``ridge''
between the two regions instead of close to the minima as in the unimodal situation.}, 
and the corresponding interpretation should be taken with caution.

\subsubsection{QT+DM}
\label{sec:QT_DM_fitting}

\begin{figure}[t]
    \centering
    \includegraphics[width=0.85\linewidth]{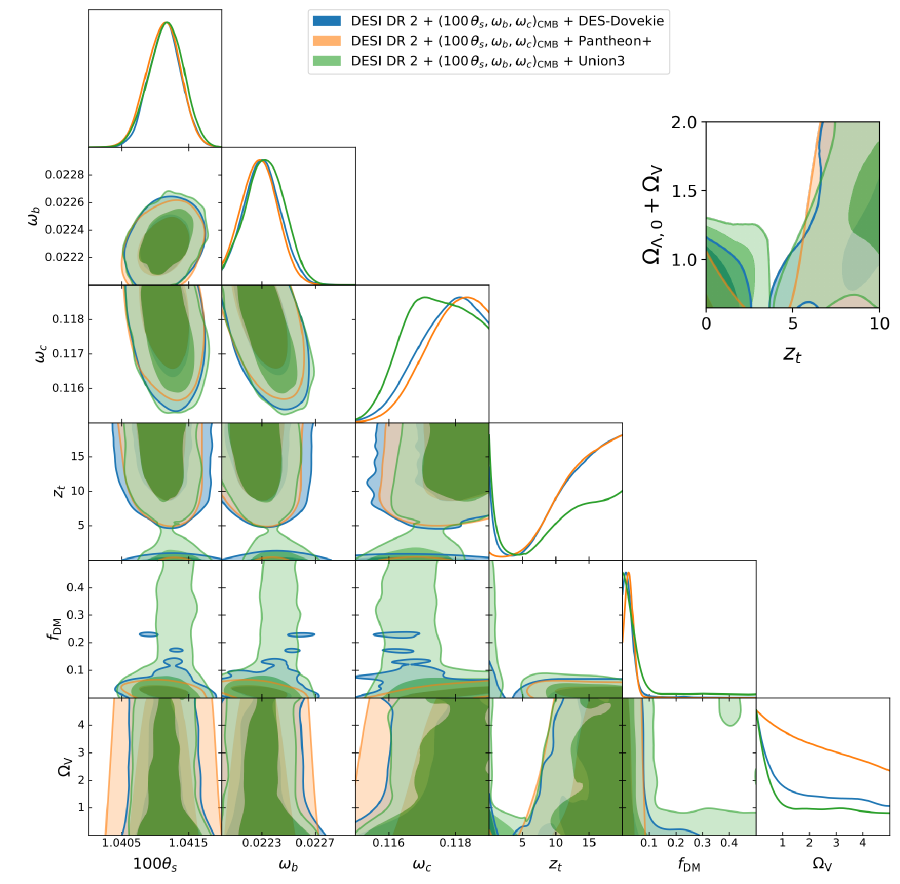}
    \caption{Same as Fig.~\ref{fig:triangle_5param}, but for the QT+DM model.}
    \label{fig:triangle_6param}
\end{figure}

In this model, only $x_\text{DW}=0$ is fixed in the full model. The posterior distributions of this model are given in Fig.~\ref{fig:triangle_6param}, and constraints and performances from different SN datasets are given in the second column from the right in Tables~\ref{tab:inference_result_DES},~\ref{tab:inference_result_Pan}, and~\ref{tab:inference_result_Uni}.

The general performance of the QT+DM model is similar to that of the QT model discussed in the previous section.
The distributions of $\zt$ and $\Omega_V$ have the same trend as the QT model and are therefore unconstrained.
However, the distribution of $\Omega_V$ exhibits a much longer tail at small $z_t$ than in the QT model.
The largeness of $\Omega_V$ is compensated by a negative $\Omega_{\Lambda,0}$, \ie, the universe at the present time has an anti-de Sitter vacuum. 
It is verified by the inset of Fig.~\ref{fig:triangle_6param} where the distribution of $\Omega_{\Lambda,0}+\Omega_V$ is shown instead of $\Omega_V$.
We have checked that the corresponding sample points have distance curves (like $D_L$) almost degenerate with those with samples of similar $\chi^2$ but with $\Omega_V<1$.
An anti-de Sitter universe can have many interesting consequences, see, \eg,~\cite{Koren:2025ymq} or~\cite{Luu:2025dax} (the latter also related to the DESI DR2 anomaly).
Further discussion along this possibility, however, is beyond the scope of this work.
Similar to the QT model, if $\zt<1$ and restricted to the dS vacuum, then 95\%(68\%) of the samples have fractional change of vacuum energy $\Omega_V/(\Omega_V+\Omega_{\Lambda,0})<0.78(0.36)$, with the median of the ratio at $\sim 0.24$.
Unlike the QT model, the QT+DM model does not face the same transition completion considerations or CMB anisotropy constraints for small $\zt$.

The new parameter $\fa$ shows a peak at $0<\fa<0.1$, and has a flat tail at large $\fa$, in particular for DES-Dovekie and Union3.
However, as seen in the $\fa$-$\zt$ joint distribution, those samples with a large $\fa$ usually also have a relatively small $\zt$, which can be intuitively understood from the perspective that it is difficult for the dataset adopted in this analysis to constrain any drastic change at $z<0.5$, as seen by the lack of low-redshift data in Fig.~\ref{fig:hubble_diag}.
Thus, if the small $\zt$ region can be somehow excluded (\eg, by CMB anisotropies), an upper bound on $\fa$ can then be inferred.
Manually excluding all samples with $\zt<1$, we may set the bound $\fa\lesssim 0.03(0.08)$ at $1\sigma$$(3\sigma)$ CL for all three datasets.
Similarly, these small-$\zt$ samples also cause the posterior distribution of $\omega_c$ to be more skewed than that for the QT model as well as the full QT+DM+DW model (Fig.~\ref{fig:triangle_7param}), in particular for Union3 (also visible for DES-Dovekie but not as strong).

Comparing this model to the others in terms of $\chi^2$, the performance of the QT+DM model is similar to the QT model, with their best-fit points almost identical for the parameters they share.\footnote{For this reason, the second-best sample is presented in Fig.~\ref{fig:hubble_diag} so that the curves are distinguishable.}
The performance of the model is therefore more penalized for containing an additional parameter, performing worse on the AIC analysis than the QT model.
On all three datasets, the Bayes factors $\ln\mc{B}$ of the QT+DM model are less than $-1.1$ (less than $-2.3$ for DES-Dovekie), indicating that this model is substantially (strongly) disfavored compared to the benchmark $\lcdm$ model according to the Jeffery scale.
The $\Delta$DIC score of the model seems plausibly good on DES-Dovekie and Union3, but not Pantheon+.
However, due to the more severe bimodality compared to the QT model, this score may not be meaningful (see footnote~\ref{fn:DIC-bimodality}).

\subsubsection{QT+DM+DW}
\label{sec:QT_DM_DW_fitting}

\begin{figure}[t!]
    \centering
    \includegraphics[width=0.95\linewidth]{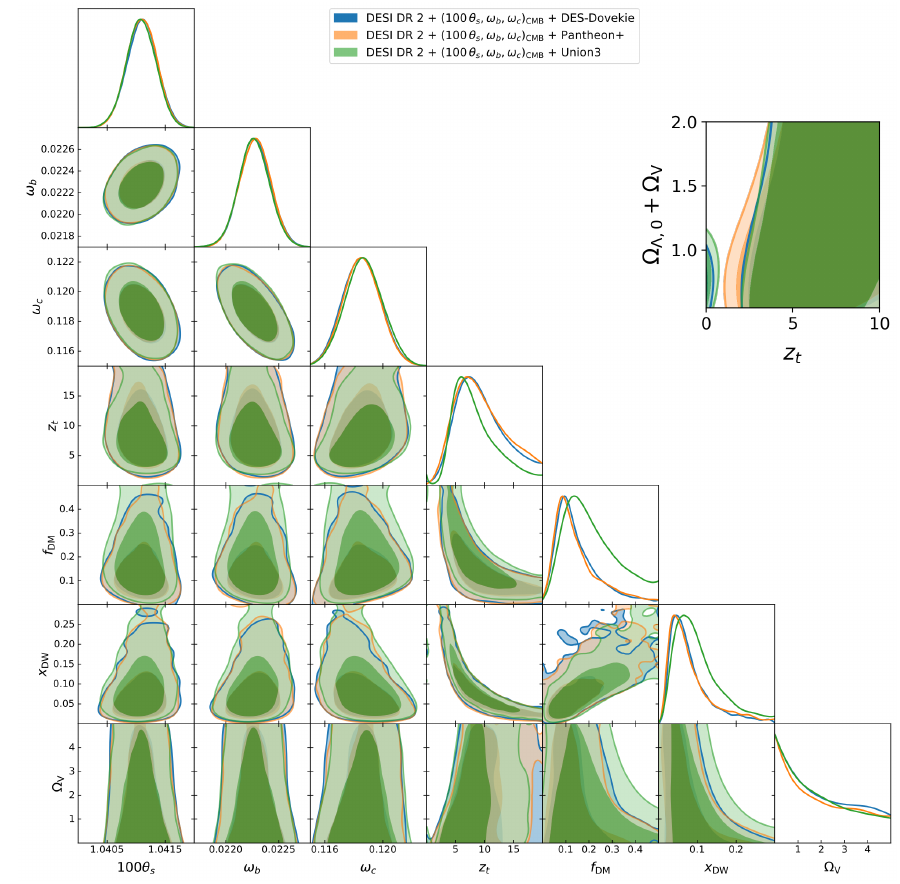}
    \caption{Same as Fig.~\ref{fig:triangle_5param}, but for the full QT+DM+DW model.}
    \label{fig:triangle_7param}
\end{figure}

The involvement of DW creation distinguishes this model from the QT and QT+DM models, as seen in Fig.~\ref{fig:triangle_7param} and the rightmost column of Tables~\ref{tab:inference_result_DES},~\ref{tab:inference_result_Pan}, and~\ref{tab:inference_result_Uni}.
The distribution of $\zt$ now has a nearly unimodal distribution, acquiring a peak at $\zt\sim 5$, although a small peak at small $\zt$ remains.
The distributions for $\fa$ and $\xdw$ are also unimodal, with the peaks at $\fa\sim 0.1$ and $\xdw\sim 0.05$, respectively.
Note that the central values of $\zt$, $\fa$, and $\xdw$ are all different from zero at the $2\sigma$ level, aside from the small peak at $z_t\sim 0$.
Such a change can be intuitively understood from the EoS of the components involved in the model.
The potential energy of the false vacuum, the massive DM before the QT, the DR after the QT, and the DWs have $w=-1,~0,~1/3$, and $-2/3$, respectively.
Without the involvement of DWs, a transition from vacuum energy and DM to DR increases $w$, giving the opposite behavior of the best-fit CPL model to the DESI DR2 results.
On the other hand, when DWs are created, they redshift more slowly and can become an important component of the total energy density, driving the universe to smaller $w$.
Interestingly, when considering a scenario where part of the DM may have an EoS deviating from $w=0$, Ref.~\cite{Braglia:2025gdo} also have the preferred range of $w$ to be around $w=-2/3$.\footnote{Ref.~\cite{Braglia:2025gdo} take the DM to be the only species with nonstandard thermal history, and is thus very different from the situation considered here. Our study also provides an underlying physics realization for such a phenomenological model.}
For the distribution of the rest of the parameters, $100\uu\theta_s$, $\omega_b$, and $\omega_c$ are well centralized around the expectation values of the compressed CMB prior, and $\Omega_V$ is again relatively flat and unconstrained as in the previous two cases.

Additionally, it can be seen from the posterior distribution that the preferred regions for $\zt$, $\xdw$, and $\fa$ exhibit a correlation. An increasing $\zt$ prefers a decreasing $\xdw$ and $\fa$. More precisely, the fit shows an approximate relation $\fa\propto (1+\zt)^{-1}$ and $\xdw \propto (1+\zt)^{-3/2}$. This is because for an earlier phase transition, the DWs have more time to increase their abundance, and the DM converted to DR has more time to decrease its abundance relative to the unconverted DM. Thus, for example, the same fraction $\xdw$ of DWs converted at an earlier $\zt$ would dominate the universe's energy budget sooner, which could put it in tension with the data for too large $\xdw$ or $\zt$. A similar argument can be made for $\fa$---too large an $\fa$ at too early a $\zt$ would modify the DM abundance by too much. On the other hand, if $\zt$ is smaller, a larger effect on $\xdw$ and $\fa$ is needed to explain the data. And if $\zt$ is too small, then the effect is not visible to the cosmological data at all.

The performance of the QT+DM+DW model is also generically better than the QT and QT+DM models considered earlier, and comparable to (if not better than) that of the CPL parametrization.
The best-fit $\chi^2$ values of the QT+DM+DW model are comparable to those of the CPL parametrization on all three SN datasets, while the AIC values are penalized by the two additional model parameters.
The Bayes factors $\ln\mc{B}$ are positive on all SN datasets, suggesting the model to be somewhat preferred, not only to the fiducial $\lcdm$ model but generically to all the other models.
For Union3, the model has $\ln\mc{B}$ around 2.3, the boundary of strong evidence, and the performance is comparable to that of the CPL parametrization.
On the other hand, for DES-Dovekie and Pantheon+, the model's $\ln\mc{B}$ has not reached the level of substantial evidence ($\ln\mc{B}>1.1$), but the model is still preferred against the CPL parametrization.
The $\Delta$DIC scores of the model suggest a strong preference on all three SN datasets.
However, the DIC interpretation should still be taken with some caution due to the small amount of bimodality in the $\zt$ distribution.

It should be noted that the QT+DM+DW model provides a fuller physical picture of the underlying physics, compared to the CPL model which is merely a parametrization.
Therefore, while the QT+DM+DW model has been statistically penalized for being more complicated, it offers a higher degree of explainability and testability.
That it still performs comparably to or better than the CPL parametrization (on all metrics besides the AIC) while generally outperforming the simpler QT and QT+DM models should lend credence to this line of model building.

\subsection{CMB anisotropy constraints}
\label{sec:ISW}

ISW-like effects on the cosmic microwave background (CMB) can result both from a phase transition itself, as well as the resulting DWs if they are allowed to form by the symmetries of the theory. First, we briefly sketch the bounds resulting from DW production. After, we provide a more rigorous treatment of the bounds on all late-universe PTs regardless of whether DWs are formed.

The existence of DWs (as well as the existence of transient nucleated bubbles) generates two effects: 1) the DWs as a source of metric perturbations that are subject to the anisotropy constraints and 2) DWs as an extra late-time energy component to change the background evolution as well as the growth of structure to affect Planck lensing reconstruction. The constraints from CMB anisotropy usually are more stringent, so we focus on this constraint.

The mean energy density of DWs with characteristic length scale $L$ is 
\beqa
\rho_{\rm DW} \sim \frac{\sigma}{L} ~.
\eeqa
The dimensionless gravitational potential at the scale of $L$ (using the Newton formula) is 
\beqa
\Phi_{L} \sim 4\pi\,G_N\,\rho_{\rm DW}\, L^2 ~. 
\eeqa
The number of domain walls in each Hubble patch is related to the DW length scale and Hubble parameter by $N^{1/3}_{\rm DW}\sim d_H/L=(LH)^{-1}$. 
For a given photon trajectory traversing the DW network, summing the contribution to the gravitational potential from each DW perpendicular to the trajectory leads to
\begin{align}
\Phi\sim\sum_i\frac{4\pi G_N\uu \rho_{\rm DW}L^3}{(i+1/2)L}\sim 4\pi G_N\uu\rho_{\rm DW}L^2\cdot\ln N_{\rm DW}\,,
\end{align}
where the Hubble patch size is used to regulate the summation. The leading $N_\text{DW}$ depedence comes from the factor of $L^2$; the logarithm induces a comparatively minor numerical correction and is therefore neglected below. Along a given line of sight, the number of DWs within a Hubble length is $\sim N_\text{DW}^{1/3}$. For the anisotropy between two different lines of sight, the expected difference in domain walls encountered is $\sim \sqrt{N_\text{DW}^{1/3}}$. Therefore, the fluctuation of the gravitational potential is
\beqa
\Phi_{\rm rms} \sim N_{\rm DW}^{1/6} \, 4\pi\,G_N\,\rho_{\rm DW}\, L^2\sim\frac{3}{2}\,\frac{\Omega_{\rm DW}(z)}{N_{\rm DW}^{1/2}}  ~,
\eeqa
where $\Omega_{\rm DW}=\rho_{\rm DW}/\rho_c$ with $\rho_c = 3 H^2/(8\pi\,G_N)$.
Frustrated DWs scale as $L \propto a$, while the Hubble distance $d_H \propto a^{3/2}$ during matter domination.
Time dependence is thus introduced into $\Phi_{\rm rms}$.
And since $a$ is generically of $\mc{O}(1)$ in the late universe, $N_\text{DW}$ can be evaluated either at the DW formation time (as it is defined throughout this work) or today, with minimal effect on the results.

This induces temperature perturbations in the CMB, $\Delta T/T \sim 2\int d \eta \, \dot{\Phi}_\text{rms}$, so 
\beqa
\frac{\Delta T}{T}\bigg|_{\rm rms} \sim \Omega_{\rm DW} \, N_{\rm DW}^{-1/2}  ~.
\eeqa
For $\ell \lesssim 30$, the CMB temperature fluctuation is dominated by the cosmic variance. The Planck measurement has $\Delta T/T\sim 1.1\times 10^{-5}$, which can be translated into a constraint on DWs as 
\beqa
\Omega_{\rm DW} \, N_{\rm DW}^{-1/2} \lesssim 10^{-5} ~. 
\eeqa

For quantum tunneling with $N_{\rm DW} = O(1)$ as in Eq.~(\ref{eq:N_DW_vacuum}), this imposes a very stringent constraint on the DW energy density fraction $\Omega_{\rm DW}$. For PTs with strongly time-dependent nucleation rates as in Eq.~(\ref{eq:N_DW}), this constraint is easily satisfied.

\subsubsection{CMB constraints on a vacuum phase transition}
\label{sec:ISW-vacuum}

Whether or not DWs form, a late phase transition can on its own induce superhorizon temperature fluctuations in the CMB.
Prior works \cite{Jinno:2017fby,Elor:2023xbz,Koren:2025ymq} have developed the formalism for calculating the anisotropy in the CMB arising from the effects of stochastic bubble nucleation on photon propagation. These works deal with the case of a thermal PT. Here, we expand on these works to also calculate the effects of a vacuum PT. This is most relevant to the QT model, so for simplicity we assume here that the vacuum energy difference is converted entirely into DR, and no coupling to DM or production of DWs exists. The following results could also be extended to the case where the DM is affected by the QT, so long as there is a negligible effect on the tunneling rate. Taking into account DM couplings or DW production would modify the expansion history, but such changes should minimally affect the results. 

\begin{figure}[t!]
    \centering
    \includegraphics[width=0.5\linewidth]{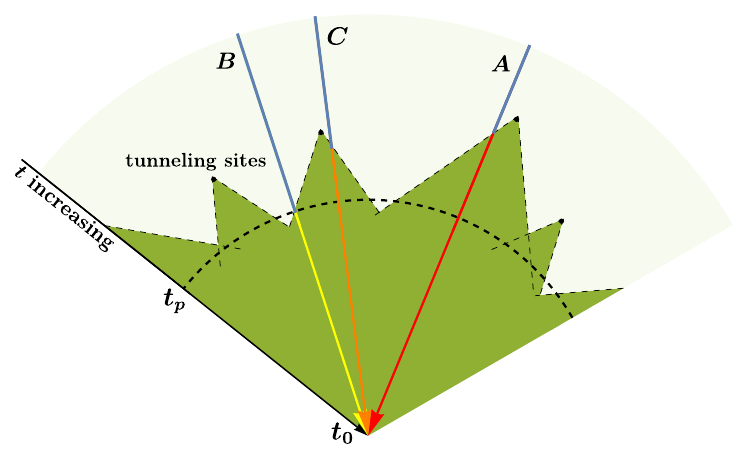}  
    \caption{Schematic illustration of the effect on CMB photons of a phase transition. Bubbles nucleate from the spacetime points denoted by black dots, filling space with the true vacuum (darker green shading). Lines A,B,C show example photon trajectories with different arrival directions. The pairing BC gives an example of a single-bubble contribution to the $\delta t$ correlation function, while AB and AC give examples of double-bubble contributions.} 
    \label{fig:ISW}
\end{figure}

For a vacuum QT with a constant nucleation rate $\gamma$, the probability for a spacetime point $x=(t_x,\mathbf{x})$ to remain in the false vacuum is
\begin{equation}
\label{eq:Psurvx}
    P_\text{surv}(x) = \exp \left[ - \int_0^{t_x} dt_n \frac{4\pi}{3} (t_x-t_n)^3 \gamma \right] = \exp \left[ - \frac{\pi}{3} t_x^4 \gamma \right] \, ,
\end{equation}
where the bubble wall velocity is assumed to be approximately the speed of light. The average transition time for a given point in space is then
\begin{equation}
    \bar{t}_c = \int_0^\infty dt_x t_x \int_0^{t_x} dt_n 4 \pi (t_x-t_n)^2 \gamma P_\text{surv}(x) = \left(\frac{3}{\pi}\right)^{1/4} \Gamma\left(\frac{5}{4}\right) \gamma^{-1/4} \, ,
\end{equation}
where $\Gamma(z)$ is the Euler gamma function. Note this is larger than the average nucleation time and smaller than the percolation time (\ref{eq:tpercolation}).

For two spacetime points $x,y$, the survival probability for both points is calculated by summing the volumes of the past light cones for each point, taking care not to double count the volume where the two past light cones overlap. 
It is given by
\begin{equation} \label{eq:Psurvxy}
    P_\text{surv} (x,y) = \exp \left[ -\mc{I}(x,y)\right]\,,
\end{equation}
with $\mc{I}$ calculated as
\begin{equation}
\label{eq:Ixy}
\begin{aligned}
\mc{I}(x,y)&=
    \gamma\cdot\left[\int_{t_\text{max}}^{t_x}dt_n \frac{4\pi}{3} (t_x-t_n)^3 + \int_0^{t_\text{max}} dt_n \frac{\pi}{3} (t_x-t_n)^3 (2+c_x) (1-c_x)^2\right] + (x \leftrightarrow y)\,,
\end{aligned}
\end{equation}
where $d \equiv |\mathbf{x}-\mathbf{y}|$, 
$c_x \equiv -\frac{d^2 + (t_x-t_n)^2 - (t_y-t_n)^2}{2 d (t_x-t_n)}$, $c_y \equiv \frac{d^2 + (t_y-t_n)^2 - (t_x-t_n)^2}{2 d (t_y-t_n)}$, and $t_\text{max} \equiv (t_x+t_y-d)/2$. The second term in the square bracket of the first line should vanish for $t_{\rm max}<0$, and the first should vanish when $t_x<0$ or $t_x<t_{\rm max}$.

With the actual transition time of the two points as $T_x$ and $T_y$, the cumulative distribution function of the two points' transition $Q(x,y)\equiv{\rm Pr}(T_x<t_x\text{ and }T_y<t_y)$ is given by
\begin{align}\begin{aligned}
Q(x,y)&=1-{\rm Pr}(T_x>t_x)-{\rm Pr}(T_y>t_y)+{\rm Pr}(T_x>t_x\text{ and }T_y>t_y)\\
&=1-P_{\rm surv}(x,y)\big\vert_{t_y=-\infty}-P_{\rm surv}(x,y)\big\vert_{t_x=-\infty}+P_{\rm surv}(x,y)\,.
\end{aligned}\end{align}
The probability distribution function of the two points' transition (\ie, the probability that the transition at $\mathbf{x}$ and $\mathbf{y}$ occurs within $(t_x, t_x+dt_x)$ and $(t_y, t_y+dt_y)$) is therefore
\begin{align}
\label{eq:pdf_2pt}
p(x,y)=\dfrac{\partial^2Q(x,y)}{\partial t_x\uu\partial t_y}=\dfrac{\partial^2P_{\rm surv}(x,y)}{\partial t_x\uu\partial t_y}=P_{\rm surv}(x,y)\left[\dfrac{\partial \mc{I}(x,y)}{\partial t_x}\dfrac{\partial \mc{I}(x,y)}{\partial t_x}-\dfrac{\partial^2 \mc{I}(x,y)}{\partial t_x\partial t_y}\right]\,,
\end{align}
where the first and second terms in the square bracket lead to the double- and single-bubble contribution identified in~\cite{Jinno:2017fby}, respectively (see Appendix~\ref{sec:bubble-calc} for further details).
Schematically, the single-bubble contribution corresponds to lines B+C in Fig.~\ref{fig:ISW}, while the double-bubble contribution corresponds to lines A+B or A+C.
The two-point function of $\delta t_c(\mathbf{x}) = t_c(\mathbf{x}) - \bar{t}_c$ is then calculated as
\begin{align}\begin{aligned}
\langle\delta t_c(\mathbf{x})\delta t_c(\mathbf{y})\rangle&=\int^\infty_0 dt_x \int^\infty_0 dt_y\,(t_x-\bar{t}_c)(t_y-\bar{t}_c)\,p(x,y)\,,
\end{aligned}\end{align}
based on which the dimensionless power spectrum for $\delta t_c$ is defined using the spherically symmetric three-dimensional Fourier transform
\begin{equation}
    P_{\delta t}(k) \equiv \frac{k^3}{2 \pi^2} H^2(t_p) \int_0^\infty dd \, 4 \pi d^2 \frac{\sin(kd)}{kd} \langle\delta t_c \delta t_c \rangle \, .
\end{equation}
Note that this uses a single value for $H=H(t_p)$, equivalent to assuming $\delta t_c \ll \bar{t}_c$.

\begin{figure}[t!]
    \centering
    \includegraphics[width=0.48\linewidth]{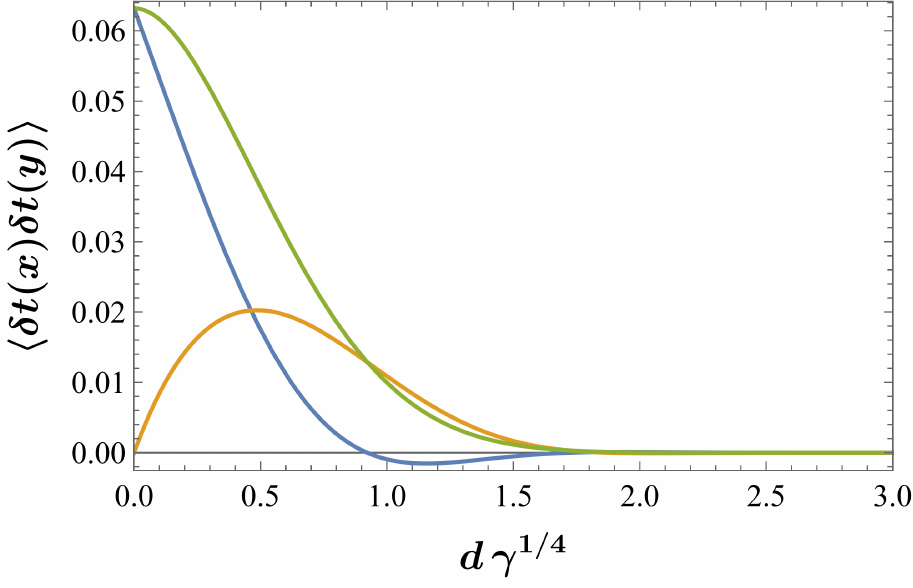}
    \hspace{3mm}
    \includegraphics[width=0.48\linewidth]{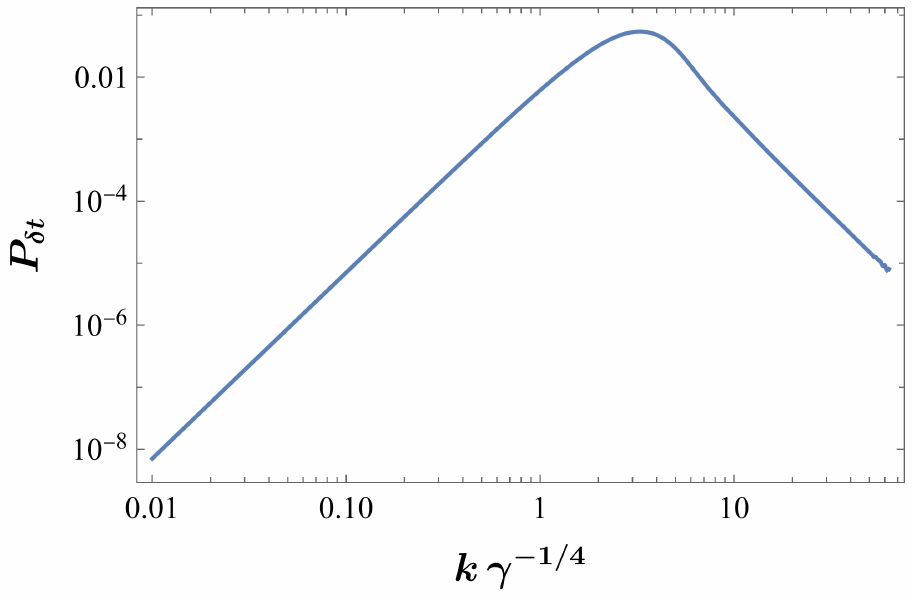}
    \caption{{\it Left:} The single-bubble (blue) and double-bubble (orange) contributions to $\langle \delta t_c (\mathbf{x}) \delta t_c (\mathbf{y}) \rangle$. Their sum is shown in green. {\it Right:} $P_{\delta t}(k)$ for the combined single- and double-bubble contributions. $H(t_p) \gamma^{-1/4}$ is taken to be unity for plotting purposes.
    }
    \label{fig:single_double_bubble}
\end{figure}

The single- and double-bubble contributions to $\langle \delta t_c (\mathbf{x}) \delta t_c (\mathbf{y}) \rangle$ are shown in the left panel of Fig.~\ref{fig:single_double_bubble}, and $P_{\delta t}(k)$ is shown in the right panel. The power spectrum can be approximated by $P_{\delta t}(k) \approx 0.007 (k \gamma^{-1/4})^3 (H(t_p) \gamma^{-1/4})^2$ for $k \ll \gamma^{1/4}$ and $P_{\delta t}(k) \approx 2 (k \gamma^{-1/4})^{-3} (H(t_p) \gamma^{-1/4})^2$ for $k \gg \gamma^{1/4}$.

The power spectrum of $\delta t_c$ needs to be converted to that of the induced photon redshift $\delta z_0$ in order to calculate the additional CMB anisotropy. As addressed in earlier sections, some fraction $r=\Omega_V/(\Omega_{\Lambda,0}+\Omega_V)$ of the total CC energy density prior to the phase transition $\Omega_\Lambda=\Omega_{\Lambda,0}+\Omega_V$ 
is assumed to convert to DR at the conclusion of the PT, while the matter energy density $\Omega_m$ is unaffected. In the limit $\delta z_0 \ll 1$, the result is
\begin{equation}
\label{eq:Pdz0}
    P_{\delta z_0} (k) \approx \frac{r^2 \Omega_\Lambda^2 \left( \Omega_\Lambda \left[1 + r \left(\frac{1}{(1+z_t)^4} - 1 \right) \right] + \Omega_m \right)}{(1+z_t)^2 (\Omega_\Lambda+ \Omega_m (1+z_t)^3)^3} k^3 \int_{k_\text{min}}^{k_\text{max}} d\kappa \, \kappa^{-1} P_{\delta t}(\kappa) \int_{-1}^1 d\mu \, s^{-3} P_{\delta t} (s) \, ,
\end{equation}
where $s = \sqrt{k^2 + \kappa^2 - 2 k \kappa \mu}$, and $k_\text{min,max}$ are chosen to ensure the peak value of $P_{\delta t}$ is numerically sampled. 
Note that this expression generalizes the results in~\cite{Koren:2025ymq}, in which $z_t \ll 1$, $rz_t \ll 1$, and $\Omega_\Lambda+\Omega_m \approx 1$ had been assumed. Its derivation is given in Appendix~\ref{app:power-spectra}.
In the following, we will approximate the redshift at which the PT completes to be the same as the redshift at percolation, $\zt \sim z_p$, although this may not always hold for vacuum PTs.

Finally, the contribution to the CMB temperature power spectrum is calculated by 
\begin{equation}
\label{eq:DlTT}
    D_\ell^{TT,\text{pt}} = 2 \ell (\ell+1) T_\text{CMB}^2 \int_{k_\text{min}}^{k_\text{max}} dk \, k^{-1} P_{\delta z_0}(k) j_\ell^2(k \Delta \tau) \, ,
\end{equation}
where $T_\text{CMB} \approx 2.73~\text{K}$ is the temperature of the CMB today, $j_\ell(x)$ is the spherical Bessel function, and $\Delta \tau = \int_0^{\bar{z}_\text{pt}} dz \, (1+z)^{-1} H_0^{-1} [\Omega_\Lambda+\Omega_m(1+z)^3]^{-1/2}$ is the comoving distance from today to the average PT redshift (under the assumption that the PT minimally impacts the expansion rate of the universe, \ie, approximating $r \approx 0$, because the full result introduces a negligible correction).

\begin{figure}[t!]
    \centering
    \includegraphics[width=0.6\linewidth]{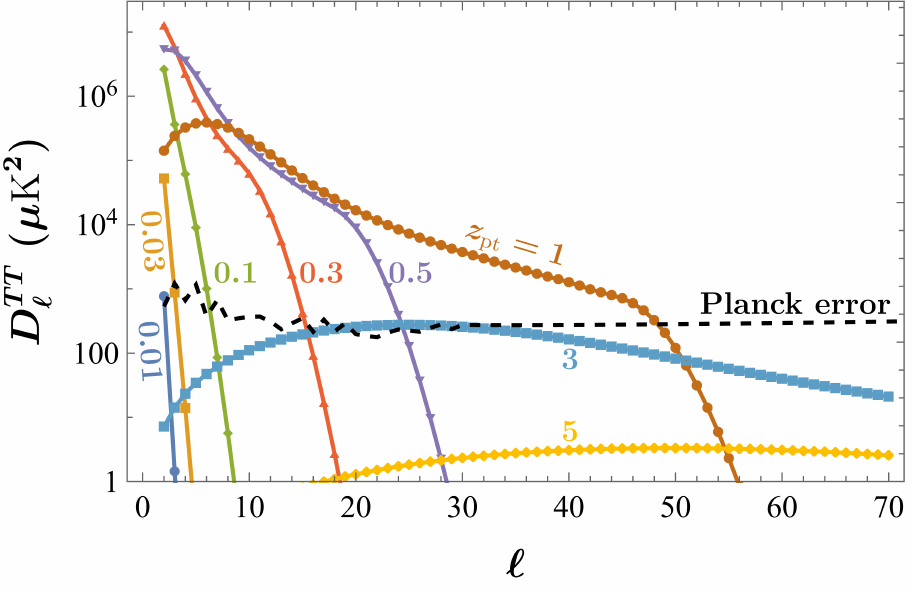}
    \caption{The CMB temperature power spectrum for various values of $\zt$, assuming $r=0.1$, $\Omega_\Lambda=0.69$, and $\Omega_m=0.31$, along with the Planck $+1\sigma$ error bar (black dashed). For the purpose of illustration, bounds on whether the PT completes are ignored here (see next figure).
    }
    \label{fig:DlTT}
\end{figure}

Example CMB multipole spectra are shown in Fig.~\ref{fig:DlTT}, along with the $1\sigma$ Planck error bars. PTs with small $\zt$ have spectra that cut off at small $\ell$ owing to the late stage and thus large scale of the perturbations. Therefore, the CMB will in general have a minimum and maximum $z_t$ for which it is capable of setting bounds.

We follow the treatment in Ref.~\cite{Koren:2025ymq} in estimating bounds. The perturbations from the phase transition induce a change in the goodness of fit approximated by
\begin{equation}
    \Delta \chi^2 \approx \sum^{\ell_p+1}_{\ell_p-1} \left(\frac{D_\ell^{TT}}{\sigma_\ell}\right)^2 \, ,
\end{equation}
where $\ell_p$ is the value of $\ell$ for which $D_\ell^{TT}$ is peaked (for given PT parameters), and $\sigma_\ell$ are the 1$\sigma$ Planck error bars \cite{PlanckData}.\footnote{When the peak value is $\ell=2$, then $\ell=2,3,4$ is used in the sum.} The 2$\sigma$ bound is given by requiring $\Delta\chi^2 \leq 5.99$. This approximation of using only a few $\ell$ bins is used to sidestep a full reanalysis of Planck data, and only a few bins are used here because the Planck error bars are highly correlated and depend on the assumed cosmology. The result is not very dependent on the exact number of $\ell$ bins used.

\begin{figure}[t]
    \centering
    \includegraphics[width=0.55\linewidth]{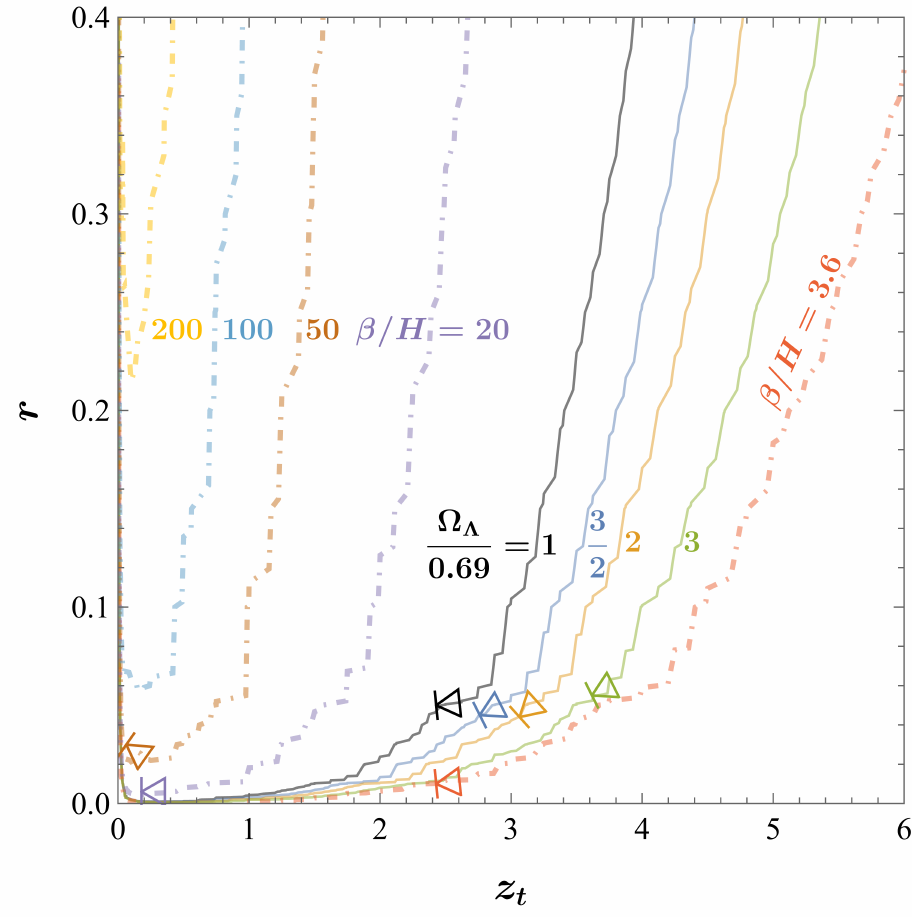}
    \caption{Bounds on $r=\Omega_V/(\Omega_{\Lambda,0}+\Omega_V)$ and $\zt$. The {\it solid} lines show the exclusion bounds for vacuum PTs with various values of $\Omega_\Lambda=\Omega_{\Lambda,0}+\Omega_V$; excluded regions are to the top-left of these lines. The {\it dot-dashed} lines show exclusion bounds for PTs with time-dependent nucleation rates with various values of $\beta/H$, fixing $\Omega_\Lambda = 0.69$. For all, $\Omega_m=0.31$. The {\it triangles terminating on lines}~($\rule{0.08ex}{1.55ex}\mkern0mu\raisebox{0.05ex}{\ensuremath{\triangleleft}}$)
    show the minimum value of $z_t$ for the color-corresponding PT to complete; these are {\it not} displayed for the $\beta/H=100,200$ curves to avoid clutter (they fall at very small $z_t$). 
    }
    \label{fig:z_r_bounds_vac}
\end{figure}

The bounds calculated in this way are shown in Fig.~\ref{fig:z_r_bounds_vac}, excluding regions above the solid lines, which correspond to various values of $\Omega_\Lambda$. 
To the left of the triangles terminating on lines~($\rule{0.08ex}{1.55ex}\mkern0mu\raisebox{0.05ex}{\ensuremath{\triangleleft}}$), 
the PT does not have time to complete before the present day, and our calculations are thus inaccurate. The minimum redshift for completion is estimated using Eq.~(\ref{eq:pt-completion}), using the same assumptions on $\Delta \tau$ as in (\ref{eq:DlTT}). 
An exploration of incomplete phase transitions is beyond the scope of this work, but it is likely that not all such incomplete phase transitions are excluded. For example, the bounds become weaker as $z_t$ approaches zero, and it is indeed expected that the CMB cannot place limits if the PT occurs at sufficiently small $z_t$.

\subsubsection{CMB constraints on a phase transition with a time-dependent nucleation rate}
\label{sec:ISW-time-dependent}

Bounds on a PT with a time-dependent nucleation rate have already been derived in Ref.~\cite{Koren:2025ymq}. Since they are similar to the calculations presented above for a vacuum PT, the formulae will not be repeated here. The main difference is that the constant $\gamma$ must be replaced by a time-dependent function $\Gamma(t) \approx \Gamma_0 e^{-S(t_f)} e^{\beta (t-t_f)}$, where $t_f$ is the approximate time of the phase transition and $S$ is the bounce action for a nucleating bubble. Bounds on the $r$-$z_t$ plane for various values of $\beta/H_*$ (fixing $\Omega_\Lambda=0.69$ and $\Omega_m=0.31$) are shown by the dot-dashed curves in Fig.~\ref{fig:z_r_bounds_vac}. For sufficiently fast PTs with $\beta/H_* \gtrsim 300$, even $r=\mathcal{O}(1)$ will not violate CMB anisotropy bounds. Such transitions are thus only subject to constraints on cosmological evolution as in Sec.~\ref{sec:DESI_fitting}.

A well-known result \cite{Enqvist:1991xw} is that the number density of bubble nucleation for a thermal PT goes as $n(t_n) = (8 \pi)^{-1} \beta^3$, with $\beta = d \ln \gamma / dt |_{t_p}$ as in Eq.~(\ref{eq:nnuc}). For a vacuum PT, $\gamma$ does not depend on time, but we could still use this relation along with (\ref{eq:Ntn_vacuum}) to define an ``effective'' $\beta_\text{eff}$ for a vacuum transition, which gives $\beta_\text{eff}/H_* = (8 \pi N(t_n))^{1/3} \approx 3.6$.
Indeed, we have verified that the $D_\ell^{TT}$ spectrum of a vacuum PT can be reasonably well approximated using the formalism of a thermal PT using $\beta/H_*=\beta_\text{eff}/H_*$ provided that $z_t \lesssim 0.5$, but it does not provide a good approximation for larger $z_t$ (due to the different time dependence of the nucleation rate for each).
This can be seen by comparing the dot-dashed red $\beta/H_*=3.6$ curve to the solid black $\Omega_\Lambda/0.69=1$ curve in Fig.~\ref{fig:z_r_bounds_vac}. (Note that these curves have the same minimum $z_t$ for the phase transition to complete, as indicated by the $\rule{0.08ex}{1.55ex}\mkern0mu\raisebox{0.05ex}{\ensuremath{\triangleleft}}$ symbols, because they have the same $N(t_n)$.)

\section{Summary and discussion}
\label{sec:discussion}

In this work, we examine the viability for a QT to occur in the late universe, from the perspective of both model building and data analysis.
On the model side, three models are built in a sequence of increasing complexity, given by (i) only a QT, (ii) both a QT and a dark matter sub-component that transitions to dark radiation following the QT (QT+DM), as well as (iii) the previous model together with domain walls generated by the QT (QT+DM+DW).
In the QT model, the potential and tunneling rate are static. The transition occurs when the quartic root of the tunneling rate is close to the Hubble parameter, leading to a small number of $\mc{O}(1-10)$ bubbles per Hubble patch.
In the QT+DM and QT+DM+DW models, the introduction of the DM coupled to the scalar field undergoing the QT allows for a DM-density-dependent term in the scalar potential. This induces a time dependence in the tunneling rate, making the transition proceed far more quickly and with many more bubble nucleation sites than the static case.

The models are fitted against the combination of the DESI DR 2 baryon acoustic oscillation data; a distilled CMB likelihood; as well as the DES-Dovekie, Pantheon+, and Union3 supernova distance datasets. CMB anisotropy constraints are shown to provide a complementary probe of these models. The QT model is generally disfavored by both cosmological expansion and CMB anisotropy constraints. It fails in explaining the tension between recent data and the $\lcdm$ model. It will not complete if the tunneling redshift is too small. When it does complete, it is constrained either to have tunneling redshift greater than of order a few, or to induce only a very small fractional change in the dark energy density. 

The QT+DM model alleviates some of these constraints because its DM-density-dependent tunneling rate allows it to complete quickly. Therefore, CMB anisotropy constraints and completion time considerations need not apply. A decrease of up to $\sim 80$\% in the total vacuum energy is allowed for a transition redshift $\zt<1$. On the other hand, it tends to perform even worse than the QT model at fitting to the joint cosmic expansion data. 

The QT+QM+DW model is much preferred compared with the other two scenarios. Like the QT+DM model, it can conclude quickly enough to avoid CMB anisotropy constraints.
It provides a much better fit to the cosmic expansion data than the $\lcdm$ model, comparable to the results of the CPL parametrization. 
An advantage of the QT+DM+DW model over the CPL parametrization is that this model contains a more complete picture of the underlying physics.
The transition in this case is preferred to occur at a redshift of around 7, with about $10\%$ of total DM involved in the transition.
It would be interesting to see if the performance of this model (as well as the DESI anomaly itself) is robust with respect to future DESI results, and if it is compatible with other cosmological observations in the near future.

For some parameter choices, the 1D posterior distributions of particular parameters remain unconstrained. This is the case for the transition redshift $\zt$ and released vacuum energy abundance $\Omega_V$ in the QT and QT+DM models. Nevertheless, the 2D joint posterior distributions for these parameters do show a correlation. Therefore, if the prior range of one parameter can be constrained, the posterior of the other can likewise be constrained. For example, if $\zt<10$ can be imposed in the QT model, Fig.~\ref{fig:triangle_5param} indicates a 2-$\sigma$ limit of $\Omega_V \lesssim 3$. There are many types of cosmological observations of inferences that could help to narrow the range of the priors. For instance, structure formation~\cite{Greene:2026gnw}, primordial black hole formation, and gravitational waves~\cite{An:2026hiq} have been used to constrain late thermal PTs. Other ideas could include 3D mapping of the 21~cm hydrogen line~\cite{Mertens:2025pvk,Munshi:2025hgk,Bowman:2018yin,Koopmans:2015sua,Singh:2021mxo} and lensing of the CMB spectrum. These subjects are left for future works.

In deriving the constraints from the CMB anisotropy, the methods employed here are only valid if the transition is completed by the present time.
However, this is a merely technical requirement arising from the approximations used in the calculation, and an unfinished transition can still influence the trajectory of CMB photons once a true vacuum bubble is encountered.
As an even smaller number of bubbles can be expected in this situation, an even larger anisotropy may be expected for the small-$\ell$ region of the CMB spectrum.
On the other hand, the smallness of the bubble number also indicates a more significant fluctuation effect, which may render some of the statistical treatment considered in this work to be invalid.
Numerical simulations may be necessary in this case, which are beyond the scope of this work but would be interesting to examine in the future.

Another possible effect on CMB anisotropy arises in the QT+DM and QT+DM+DW models, where the tunneling rate depends on the DM density. Because of the density dependence, bubbles may nucleate sooner in regions of low DM density (voids) than in regions of high DM density (clusters). This may produce additional large-scale inhomogeneities correlated with large scale structures, which would be interesting to explore in future works.

\subsubsection*{Acknowledgments}
The work of YB is supported by the U.S. Department of Energy under the contract DE-SC0017647 and DE-AC02-06CH11357 at Argonne National Laboratory. The work of SL is supported by the National Science Foundation of China under Grant No.~12505128. The work of NO is partially supported by the National Science Centre, Poland, under research grant no.~2020/38/E/ST2/00243. We thank the Center for High Throughput Computing at the University of Wisconsin-Madison for providing computing resources~\cite{https://doi.org/10.21231/gnt1-hw21}.

\appendix

\section{Binning and visualization of the SN data}
\label{app:binning}

Due to the size of the SN datasets, the measurements are re-binned into seven redshift bins.
We choose the bin edges to be $0$, $0.12$, $0.24$, $0.33$, $0.42$, $0.61$, $0.81$, $100$.
Note that the binning scheme is different from that of DESI DR2, and the binning results are therefore different from DESI DR2~\cite{DESI:2025zgx}.
As the fiducial SN absolute magnitude is unknown, the distance moduli curves have the freedom to be shifted by a constant $\delta\mu$.
Thus, for a fair comparison, we shift the model curves such that they have the same weighted mean as the dataset, following the treatment of DESI DR2~\cite{DESI:2025zgx}.
Specifically, let $(z_i, \mu_i)$ be the SN distance moduli measurements, $C$ the covariance matrix of $\mu_i$, and $\hat{\mu}(z)$ be a model, we will add a shift $\delta\hat{\mu}$ to the model such that
\begin{align}
\dfrac{\mathbf{1}^T\cdot C^{-1}\cdot(\vec{\mu}-\vec{\hat{\mu}}-\delta\hat{\mu}\mathbf{1})}{\mathbf{1}^T\cdot C^{-1}\cdot\mathbf{1}}=0\,,
\end{align}
where $\vec{\mu}=(\mu_i)$ and $\vec{\hat{\mu}}=(\hat{\mu}(z_i))$ represent the measurements and model predictions at the corresponding redshifts.
The immediate result of this shift is to remove the common deviation from the model among the measurements such that the size of the error bars are shrunk~\cite{Rubin:2023jdq}.

With this shift, the re-binning is performed in the following way.
Let $H=(H_{i\alpha})$ be the projection matrix such that
\begin{align}
H_{i\alpha}=\begin{cases}
1 & \mu_i\in \text{bin }\alpha\\[1mm]
0 & \text{otherwise}
\end{cases}\,.
\end{align}
The re-binning matrix $B$ can be correspondingly defined as
\begin{align}
B=\left(H^TC^{-1}H\right)^{-1}H^TC^{-1}\,.
\end{align}
The covariance matrix after the re-binning is
\begin{align}
\widetilde{C}=B\uu C B^T-\dfrac{(B\cdot\mathbf{1})(B\cdot\mathbf{1})^T}{\mathbf{1}^T\cdot C^{-1}\cdot\mathbf{1}}\,,
\end{align}
where the second term is responsible for the freedom of the constant shift.

After this shift and re-binning, the presented distance moduli difference in the right column of Fig.~\ref{fig:hubble_diag} is $B(\vec{\mu}-\vec{\hat{\mu}}-\delta\hat{\mu}\mathbf{1})$, at redshift $B\,\vec{z}$.
The error bar sizes are chosen to be the square root of the corresponding diagonal element in $\widetilde{C}$.

For completeness we also provide in Table~\ref{tab:benchmark_curves} the model parameters for the curves that are shown in Fig.~\ref{fig:hubble_diag}.

\begin{table}[]
\renewcommand{\arraystretch}{1.2}
\centering
{\footnotesize
\begin{tabular}{c|c|c|c|c|c|c|c}
\hline
& $\theta_s$ & $\omega_b$ & $\omega_c$ & $\zt$ & $\fa$ & $\xdw$ & $\Omega_V$ \\
\hline
DES, QT & 
1.041285& 0.022412& 0.116717& 0.064305& -& -& 0.265918\\
Pan, QT &
1.041312& 0.022434& 0.116424& 0.449255& -& -& 0.076005\\
Uni, QT &
1.041298& 0.022422& 0.116577& 0.164890& -& -& 0.214735\\
\hline
DES, QT+DM&
1.0413007& 0.022435& 0.116597& 0.065569& 0.056467& -&  0.263682\\
Pan, QT+DM&
1.0412289& 0.022387& 0.116830& 0.012309& 0.127802& -&  2.371604\\
Uni, QT+DM&
1.0413087& 0.022439& 0.116545& 0.153282& 0.153108& -& 0.227245\\
\hline
DES, QT+DM+DW&
1.041038&  0.022234&  0.119138&  9.155338&  0.125846&  0.069196&      0.0\\
Pan, QT+DM+DW&
1.041035&  0.022232&  0.119163&  10.328743&  0.110694&  0.057845&  0.040837\\
Uni, QT+DM+DW&
1.041117&  0.022281&  0.118438&  2.920176&      0.5&     0.3&  0.231888\\
\hline
\end{tabular}
}
\caption{Model parameters for the benchmark curves shown in Fig.~\ref{fig:hubble_diag}. DES, Pan, and Uni are short for DES-Dovekie, Pantheon+, and Union3. 
}
\label{tab:benchmark_curves}
\end{table}

\section{Alternate formulation of the transition time correlation function}
\label{sec:bubble-calc}

The notation introduced in Sec.~\ref{sec:ISW-vacuum}, in particular Eq.~(\ref{eq:pdf_2pt}), provides a straightforward mathematical interpretation of the single- and double-bubble contributions to $\langle \delta t_c (\mathbf{x}) \delta t_c (\mathbf{y}) \rangle$. For completeness and ease of comparison, this Appendix provides expressions matching the notation in Refs.~\cite{Jinno:2017fby,Elor:2023xbz,Koren:2025ymq}, which gives a more geometric interpretation using past light cones of spacetime points.

First, the expression in Eq.~(\ref{eq:Ixy}) used for $P_\text{surv}(x,y)$ is evaluated as
\begin{equation} \label{eq:Ixy_2}
\begin{aligned}
    \mathcal{I} (x,y) 
    =& \frac{\pi \gamma}{48 d} \left[ d^5 - 4 d^4 \txpy - 2 d^3 \txmy^2 + 4 d^2 \left( 3 \txmy^2 \txpy + 4 \txpy^3 \right) \right.
    \\
    & \;\;\;\;\;\;\; \left.
    + 2 d \left( \txmy^4  + 12 \txmy^2 \txpy^2  + 8 \txpy^4 \right) + 16 \txmy^2 \txpy^3   \right] 
\end{aligned}
\end{equation}
Here, $\txpy \equiv (t_x+t_y)/2$ and $\txmy \equiv t_x-t_y$.
It has been assumed in this expression for $P_\text{surv}(x,y)$ that the past light cones of $x$ and $y$ do overlap, meaning $t_\text{max}>0$, and that neither point is inside the past light cone of the other point, meaning $|\txmy| < d$. The survival probability in these other regions are related trivially to the single-point survival probability in Eq.~(\ref{eq:Psurvx}).

The next step is to calculate the two-point correlation function of $\delta t_c(\mathbf{x}) = t_c(\mathbf{x}) - \bar{t}_c$, the difference in the actual and average vacuum transition time for a point $\mathbf{x}$. There are two contributions to the correlation function which must be added together. First, is the single-bubble case where the two points $\mathbf{x},\mathbf{y}$ are converted by the same nucleation bubble (as for lines B+C in Fig.~\ref{fig:ISW}). It is given by
\begin{equation}
\begin{aligned}
    \langle \delta t_c (\mathbf{x}) \delta t_c (\mathbf{y}) \rangle^{(s)} & = \int_{d/2}^{\infty} d \txpy \int_{-d}^{d} d \txmy \int_{d/2}^{\txpy} d \txyn   \frac{2 \pi \gamma}{d} P_\text{surv}(x,y) \left( \txpy + \frac{\txmy}{2} - \bar{t}_c \right) 
    \\
    & \qquad \qquad \qquad \quad  \times \left( \txpy - \frac{\txmy}{2} - \bar{t}_c \right)  \left( \txyn + \frac{\txmy}{2} \right) \left( \txyn - \frac{\txmy}{2} \right) \, ,
\end{aligned}
\end{equation}
where $\txyn \equiv \txpy - t_n$. The $\txyn$ integral can be performed analytically, and the other two integrations can be performed numerically. The $\txpy$ upper integration bound is approximated as infinity because $P_\text{surv}$ should anyways go to zero at large $\txpy$. 

The other contribution to the correlation function is the double-bubble case where the two points are converted by different nucleation bubbles (as for lines A+B or A+C in Fig.~\ref{fig:ISW}). It is
\begin{equation}
\begin{aligned}
    \langle \delta t_c (\mathbf{x}) \delta t_c (\mathbf{y}) \rangle^{(d)} = 
    \int_{-d}^{d} & d \txmy \int_{|\txmy/2|}^{\infty} d \txpy \int_{0}^{\txpy+\txmy/2} d t_{x,xn} \int_{0}^{\txpy-\txmy/2} d t_{y,yn}  
    \\
    & \times 16 \pi^2 \gamma^2 P_\text{surv}(x,y) \left( \txpy + \frac{\txmy}{2} - \bar{t}_c \right) \left( \txpy - \frac{\txmy}{2} - \bar{t}_c \right) t_{x,xn}^2 t_{y,yn}^2 f_x f_y \, ,
\end{aligned}
\end{equation}
where $t_{x,xn} \equiv t_x - t_{xn}$ (and similarly for $y$). Note these are integrated from zero rather than $d/2$ as in the single-bubble case (in the single-bubble case, the lower bound comes from the requirement that the same nucleated bubble needs to reach both spacetime points). The lower bound on the $\txpy$ integration comes from the requirement $t_x>0$ and $t_y>0$. In the region $|\txmy/2|<\txpy<d/2$, $P_\text{surv}(x,y)=P_\text{surv}(x) P_\text{surv}(y)$ rather than the expression in (\ref{eq:Psurvxy}). The $f$ functions, which account for the nontrivial geometry of the overlap of the past light cones of $x$ and $y$, are given by
\begin{align}
    f_x &= \left\{ \begin{array}{l l}
\frac{(t_{x,xn}+d)^2-(t_{x,xn}-\txmy)^2}{4 d t_{x,xn}} ~\, ~~ & \text{for} ~ t_{x,xn}>\frac{1}{2}(\txmy+d)  ~,
\\
1 ~\, ~~ & \text{otherwise} ~.
\end{array} \right.
    \\
    f_y & = \left\{ \begin{array}{l l}
\frac{(t_{y,yn}+d)^2-(t_{y,yn}+\txmy)^2}{4 d t_{y,yn}} ~\, ~~ & \text{for} ~ t_{y,yn}>\frac{1}{2}(-\txmy+d)  ~,
\\
1 ~\, ~~ & \text{otherwise} ~.
\end{array} \right.
\end{align}
Similar to the single-bubble case, the $t_{x,xn}$ and $t_{y,yn}$ integrals can be performed analytically, and the other two integrations can be performed numerically. Note that in the region $|\txmy/2|<\txpy<d/2$, $f_x=f_y=1$.

\section{Relationship between $\delta t$ and $\delta z_0$}
\label{app:power-spectra}

Let us derive the relationship between $\delta t$ and $\delta z_0$ used in (\ref{eq:Pdz0}). We follow the treatment of \cite{Koren:2025ymq} but impose fewer assumptions. Following their appendix, we have the following relation requiring that two different photons with the same initial redshift must travel the same comoving distance, regardless of whether they encounter fluctuations in the tunneling redshift:
\begin{equation}
\label{eq:comoving-dist-relation}
\begin{aligned}
    & \int_0^{\delta z_0} \frac{dz}{(1+z) \sqrt{(1-r) \Omega_\Lambda + r \Omega_\Lambda\left(\frac{1+z}{1+z_t}\right)^4 + \Omega_m (1+z)^3} } 
    \\
    & = \int_{z_t}^{z_t + \delta z_t} dz \left[\frac{1}{(1+z) \sqrt{(1-r) \Omega_\Lambda + r \Omega_\Lambda\left(\frac{1+z}{1+z_t+\delta z_t}\right)^4 + \Omega_m (1+z)^3}} - \frac{1}{(1+z)\sqrt{\Omega_\Lambda + \Omega_m(1+z)^3}} \right] \, .
\end{aligned}
\end{equation}
The integral on the left side of (\ref{eq:comoving-dist-relation}) can be approximated by taking the denominator as approximately constant for small $\delta z_0 \ll 1$ and substituting $z=0$. Therefore, this integral becomes
\begin{equation}
\label{eq:app-lhs}
    \approx \frac{\delta z_0}{\sqrt{\Omega_\Lambda \left[1 + r \left(\frac{1}{(1+z_t)^4} - 1 \right) \right] + \Omega_m}} \, .
\end{equation}

The integral on the right side of (\ref{eq:comoving-dist-relation}) can be approximated by using
\begin{equation}
\label{eq:dz0-RHS-binomial}
    \left(\frac{1+z}{1+z_t+\delta z_t}\right)^4 = \left(\frac{1+z}{1+z+(z_t+\delta z_t-z)}\right)^4 \approx 1 - 4 \frac{z_t+\delta z_t-z}{1+z} \, ,
\end{equation}
where we have used the integral range $z_t < z < z_t+\delta z_t$ with $\delta z_t \ll 1$ to apply the binomial approximation. Therefore,
\begin{equation}
\begin{aligned}
    & \sqrt{(1-r) \Omega_\Lambda + r \Omega_\Lambda\left(\frac{1+z}{1+z_t+\delta z_t}\right)^4 + \Omega_m (1+z)^3} 
    \\
    & \approx \sqrt{\Omega_\Lambda + \Omega_m (1+z)^3} \left[1 + 2 r \Omega_\Lambda  \frac{z_t+\delta z_t-z}{(1+z)(\Omega_\Lambda + \Omega_m (1+z)^3)} \right]^{-1}
\end{aligned}
\end{equation}
This leads to a partial cancellation of the two terms on the right side of (\ref{eq:comoving-dist-relation}), leaving behind the integral on the right side as
\begin{equation}
\label{eq:app-rhs}
    \approx \int_{z_t}^{z_t+\delta z_t} dz \, 2 r \Omega_\Lambda  \frac{z_t+\delta z_t-z}{(1+z)^2(\Omega_\Lambda + \Omega_m (1+z)^3)^{3/2}}
    \approx \frac{r \Omega_\Lambda \delta z_t^2}{(1+z_t)^2(\Omega_\Lambda + \Omega_m (1+z_t)^3)^{3/2}} \, ,
\end{equation}
where as before the denominator is taken to be approximately constant over the integration range.

Putting the two sides, (\ref{eq:app-lhs}) and (\ref{eq:app-rhs}), together,
\begin{equation}
    \delta z_0 \approx \delta z_t^2 \times \frac{r \Omega_\Lambda\sqrt{\Omega_\Lambda \left[1 + r \left(\frac{1}{(1+z_t)^4} - 1 \right) \right] + \Omega_m}}{(1+z_t)^2(\Omega_\Lambda + \Omega_m (1+z_t)^3)^{3/2}} \, .
\end{equation}
This equation differs from \cite{Koren:2025ymq} in two ways: i) they approximate the square root in the numerator as unity by taking $z_t \ll 1$, $r z_t \ll 1$, and $\Omega_\Lambda+\Omega_m \approx 1$; and ii) they have only one factor of $1+z_t$ rather than two in the denominator because in the approximation in (\ref{eq:dz0-RHS-binomial}), they appear to have simply taken $z_t \ll 1$ so that $(1+z)(1+z_t+\delta z_t)^{-1} \approx (1+z-z_t-\delta z_t)$.

The rest of the relationship between $\delta t$ and $\delta z_0$ follows the results of \cite{Koren:2025ymq} exactly. Using $\delta a_t / a_t = - \delta z_t / (1+z_t) = H_* \delta t$ gives the relationship $\mathcal{P}_{\delta z_t}(k) \approx (1+z_t)^2 \mathcal{P}_{\delta t} (k)$. The full derivation of the relationship between $\mathcal{P}_{\delta z_0}$ and $\mathcal{P}_{\delta z_t}$ can be found in \cite{Koren:2025ymq}; for our purposes only the prefactor changes as outlined above.

\bibliography{DW_DESI}
\bibliographystyle{JHEP}

\end{document}